\documentclass[12pt]{elsarticle}

\pdfoutput=1

\usepackage[top=2cm,bottom=2.65cm,left=2.65cm,right=2.55cm]{geometry}
\usepackage{graphicx} 
\usepackage{amssymb,amsmath,amsthm}
\usepackage{latexsym}
\usepackage{fullpage}
\usepackage{cases}
\usepackage{float} 
\usepackage{subfigure}
\usepackage{xargs}
\usepackage{hyperref,pdfsync}
\usepackage{ifthen}
\usepackage{tikz}
\usepackage{color}
\usepackage{ulem}
\usepackage{dsfont}
\usepackage{algorithm}
\usepackage{algpseudocode}
\usepackage{amsmath}
\usepackage{afterpage}
\usepackage{epstopdf}
\usepackage{enumerate}
\usepackage{textcomp}
\usepackage{tikz}
\usepackage{multirow} 
\usepackage[textwidth=4cm, textsize=footnotesize]{todonotes}
\usepackage[latin1]{inputenc}
\usepackage{aliascnt}

\newaliascnt{rem}{thm}
\newtheorem{rem}[rem]{Remark}
\aliascntresetthe{rem}

\newcommand*{\AdjustMargins}{%
    \setlength{\beamer@rightmargin}{0em}%
    \setlength{\beamer@leftmargin}{0em}%
}

\newcommand{\thickhline}{%
        \noalign {\ifnum 0=`}\fi \hrule height 1pt
        \futurelet \reserved@a \@xhline
}	
		
\def\1{\mathds{1}}

\def\alphaSet{\mathcal{A}}

\def\balpha{\boldsymbol{\alpha}}

\def\bomega{\boldsymbol{\omega}}
\def\bU{\tilde{\bX}}
\def\bu{\tilde{\bx}}
\def\bW{\mathbf{W}}
\def\bX{\mathbf{X}}
\def\bx{\mathbf{x}}
\def\bXs{\bX^{\star}}
\def\btY{\mathbf{\tilde{Y}}}
\def\bY{\mathbf{Y}}

\def\bc{\mathbf{c}}

\newcommand{\cluster}[1]{\mathcal{C}_{#1}}
\def\CC{\texttt{CC}}

\def\Dset{\mathbb{D}}
\def\defo{\beta}
\def\defoSpace{\mathbb{B}}
\def\design{\Omega}
\def\dimDefo{d_{\defo}}
\def\dimF{m}
\def\dimLoc{d_{V}}

\def\dimX{d_{\bX}}
\def\dimY{|\design|}

\def\eqsp{\;}
\def\eg{\textit{e.g.}\,}
\def\esp{\mathbb{E}}
\newcommand{\estim}[1]{\hat{\param}_{#1}}

\def\Fset{\mathbb{F}}

\def\gam{\mathrm{Gamma}}
\def\gammaFct{\mathcal{G}}

\def\ie{\textit{i.e.}\,}
\def\Iset{\mathbb{I}}
\def\iid{\textit{i.i.d.}}

\newcommand{\K}[1]{K_{#1}}

\def\loc{\delta}
\def\like{\text{L}}
\newcommand{\local}[2]{V(#1,#2)}

\def\multinomial{\text{Multi}}

\def\nset{\mathbb{N}}

\def\norm{\mathcal{N}}

\newcommand{\ooint}[2]{(#1,#2)}

\def\param{\theta}
\def\paramSet{\Theta}
\newcommand{\posDefMat}[1]{{\mathcal{M}^{+}_{#1}}(\rset)}
\newcommand{\pscal} [3]{\left\langle #2 , #3\right\rangle_{#1}}

\def\rmd{\mathrm{d}}
\def\rset{\mathbb{R}}
\newcommand{\rigid}[2]{T(#1,#2)}
\newcommand{\rotation}[1]{\mathcal{R}_{#1}}

\def\sca{\lambda}

\newcommandx{\tK}[1]{\tilde{K}_{#1}}

\newcommandx{\target}[3]{
\ifthenelse{\equal{#2}{}}{\pi_{#1}(\,\cdot\,|\,#3)}{\pi_{#1}(#2\,|\,#3)}
}
\newcommandx{\targetCC}[3]{
\ifthenelse{\equal{#2}{}}{\tilde{\pi}_{#1}(\,\cdot\,|\,#3)}{\tilde{\pi}_{#1}(#2\,|\,#3)}
}

\def\ups{\upsilon}
\def\Uset{\mathbb{U}}

\def\wrt{with respect to}

\def\Xset{\mathbb{X}}

\allowdisplaybreaks

\journal{}
\begin{document}
\begin{frontmatter}

\title{Online EM for Functional Data}

\author[ucd]{Florian Maire\corref{cor1}}
\ead{florian.maire@ucd.ie}
\author[telecom]{Eric Moulines}
\author[onera]{Sidonie Lefebvre}
\cortext[cor1]{Corresponding author}
\address[ucd]{School of Mathematics and Statistics, University College Dublin, Ireland}
\address[telecom]{CMAP, {\'E}cole Polytechnique, 91128 Palaiseau, France }
\address[onera]{ONERA - the French Aerospace Lab,  F-91761 Palaiseau, France}

\begin{abstract}
A novel approach to perform unsupervised sequential learning for functional data is proposed. Our goal is to extract reference shapes (referred to as \textit{templates})
from noisy, deformed and censored realizations of curves and images. Our model generalizes the Bayesian dense deformable template model \cite{Allassonniere:Mixture,Allassonniere:survey}, a hierarchical model in which the template is the function to be estimated and the deformation is a nuisance, assumed to be random with a known prior distribution.  The templates are estimated using a Monte Carlo version of the online Expectation-Maximization (EM) algorithm, extending~\cite{Cappe:Online}. Our sequential inference framework is significantly more computationally efficient than equivalent batch learning algorithms, especially when the missing data is high-dimensional. Some numerical illustrations on curve registration problem and templates extraction from images are provided to support our findings.
\end{abstract}

\end{frontmatter}

\section{Introduction}\label{sec:intro}

Functional data analysis is concerned with the analysis of curves and shapes, which often display common patterns but also variations (in amplitude, orientations, time-space warping, etc...). The problem of extracting common patterns (referred to as \textit{templates}) from functional data,
and the related problem of curves/images registration has given raised to a wealth of research efforts;~ see \cite{ramsay:functional,zhong:2008,ramsay:2011} and the references therein.

Most of the proposed techniques used so far have been developed in a supervised classification context. The method typically aims at finding a time/space warping transformation allowing to synchronize/register all the observations associated to a given class of curves/shapes and to estimate a template by computing a cross-sectional mean of the aligned patterns. In most cases, the deformation is penalized, to favor "small" time/space shifts. Many different deformation models have been proposed for curves and for images. For curves, the warping function is often assumed to be monotone increasing. In this context, the dynamic time warping algorithm is by far the most popular algorithm: it enables the alignment of curves by minimizing a cost function of the warping path, which can be solved by a dynamic programming algorithm \cite{Wang:warpingTime}. Non parametric \cite{kneip:statistical,Silvermann:nonParam,Ramsay:curveReg} as well as Bayesian approaches \cite{Telesca:bayesCurRe,Liu:Simultaneous} have also been proposed, but they are still far less popular. The situation is more complex for shapes and images. Different deformation models  have been proposed, involving rigid deformations, small deformations~\cite{Castellanos:localWarp} or deformation fields ruled by a differential equation; see \cite{Christensen:elas_trans}.

In this paper, we introduce a common Bayesian statistical framework for \textit{unsupervised} clustering and template extraction, with applications to curve synchronization and shape registration. Following the seminal work by \cite{Allassonniere:Mixture} and \cite{Allassonniere:Toward}, we generalize the mixture of \textit{deformable template models}. This approach models a curve/shape as a template (defined as a function of time or space), selected from a collection of templates, which undergoes a random deformation and is observed in presence of an additive noise; see \cite{Allassonniere:Toward,Bigot:defTemp,Christensen:defTemp} and \cite{Allassonniere:survey} for a complete survey.  Contrary to the classical time warping/spatial registration algorithms which consists in synchronizing all the observations of a shape in a supervised framework, the mixture of deformable template models is an unsupervised classifier: it estimates functional templates from a set of shapes/curves and consider the time warping/spatial deformations as a random nuisance parameter. It is important to stress that the model allows to integrate the deformation conditionally on the observations while considering the templates as unknown deterministic functional parameters. In this context, the deformation might be seen as a \textit{random effect}, which is similar to random effects in linear mixed models in longitudinal data analysis. Whereas this change in perspective might seem rather benign, it makes a huge difference both in theory and in practice.

In our model, the warping/deformation function and the cluster index is modeled as hidden data and we consequently turn to an Expectation-Maximization (EM)-type algorithm \cite{Dempster:EM} to estimate the templates. However, in our model the conditional expectation of the complete data log-likelihood is analytically intractable, compromising a plain EM implementation. This situation has raised a significant research interest over the last decades and several versions of so-called stochastic EM, in which the E-step is approximated, have been successfully applied to the template extraction problem. A rough approximation of the conditional expectation was considered in \cite{ma:bayesian}, in which the posterior distribution is replaced by a point mass located at the posterior mode. Another elementary approach consists in linearizing the deformed template in the neighborhood of its nominal shape, under the assumption of small deformations. This alternative has been considered, among others by \cite{Liu:Simultaneous} and \cite{Frey:EM_clust}, in which the transformed mixture of Gaussian models was used. Another way to handle the E-step, proposed by \cite{Gaffney:curCluAli}, consists in performing an approximate Bayesian integration, which amounts to replace the posterior distribution of the hidden data conditionally to the observation by a Gaussian distribution, obtained from a Laplace approximation. Here again, it is not always easy to justify such approximations. The expectation can also be approximated by Markov chain Monte Carlo, an idea which was put forward by \cite{Allassonniere:Mixture} and \cite{Kuhn:MCMCSAEM}, extending the original Stochastic Approximation EM (SAEM) \cite{Delyon:ConvSAEM} and known as the MCMC-SAEM algorithm. This algorithm has been theoretically justified \cite{Kuhn:MCMCSAEM} and has shown to perform satisfactorily in the template extraction application \cite{Allassonniere:Mixture}. However it turns out to be a time-consuming solution especially when a large number of observations are available and the dimension of the missing data is huge. The extension of the model to multiple classes is even more computationally involved.

We propose the Monte Carlo online EM (MCoEM), an online algorithm in which the curves/shapes are processed one at a time and only once, allowing to estimate the unknown parameters of the mixture of deformable templates model. We adapt the online EM algorithm proposed in \cite{Cappe:Online} to intractable E-step settings whereby casting MCoEM as a \textit{noisy} online EM. Our model is too general to allow the linearization or the use of Gaussian approximation of the complete data log-likelihood, as it was done in \cite{Liu:Simultaneous} and \cite{Gaffney:curCluAli}. We thus propose to approximate the conditional expectation thanks to an MCMC algorithm adapted from the celebrated Carlin and Chib algorithm \cite{Carlin:Bayesian}. Indeed, working online implies processing the data on the fly without storing them afterwards and this requires the posterior distribution exploration to be more accurate than in the MCMC-SAEM framework which refines the state-space exploration gradually, at each EM iteration. Building an online learning framework for template extraction has a two-fold motivation: (i) the data need not be stored which can be useful should the algorithm be implemented on a portable device with limited memory/energy resources and (ii) MCoEM reduces significantly the computational burden that would be generated by an equivalent batch algorithm such as the MCMC-SAEM \cite{Kuhn:MCMCSAEM}.

This paper is organized as follows: in Section \ref{sec:model} the mixture of the dense deformable template model is generalized and the Monte Carlo online EM algorithm is presented in Section \ref{sec:onlineEM}. The sampling method of the joint posterior distribution is proposed in Section \ref{sec:carlinChib}. Illustrations of templates obtained by applying MCoEM to curves and shapes are proposed in Section \ref{sec:illustration} and are compared with those obtained using MCMC-SAEM. An application of the methodology to a classification problem is provided in Section \ref{sec:classification} and shows how competitive MCoEM is over batch equivalent algorithms. Benefits and shortcomings of the MCoEM methodology are discussed in Section \ref{sec:conclusion} and perspectives raised.

\section{A mixture of deformable template model}
\label{sec:model}
\subsection{A basic deformable model}
In this section, we introduce a basic model for curves and images. A \textit{template} is a function defined on a  space $\Uset$ and taking for simplicity real values. Typically, for curves $\Uset=\rset$ and for shapes $\Uset=\rset^{2}$. We denote by $\Fset$ the set of templates.

The observations are modeled as the stochastic process $Y$ indexed by $u\in\Uset$ and given by:
\begin{equation}
\label{eq:modelFunc}
Y(u)=\sca\,f\circ D(u,\defo) + \sigma W(u)\eqsp,
\end{equation}
where, $f\in\Fset$ is a template function, $\sca\in\rset^{+\,\ast}$ is a scaling factor, $\sigma^{2}\in\rset^{+\,\ast}$ is the noise variance and $W$ a Gaussian process with zero-mean, unit variance and known covariance function. $D$ is a function, belonging to $\Dset$, the set of mappings from $\Uset$ to itself and parameterized by a vector $\defo\in\defoSpace$, where $\defoSpace$ is an open subset of some euclidean space of dimension $\dimDefo$. For curves, $\Dset$ can be chosen as the homotheties and translations mappings and more generally as the set of monotone functions (with appropriate smoothness conditions). For shapes, $\Dset$ can be taken as the set of rigid transformations of the plane, such as rotations, homotheties or translations and a local deformation field. The models for the set of deformations $\Dset$ are problem dependent; see Section \ref{sec:illustration}.

In this setting, $\defo$ and $\sca$ are random variables and  each realization of $Y$ follows from different realizations of $\defo$ and $\sca$. The quantity of interest is the template $f$ (a deterministic functional parameter), while the deformation $D$ and the global scaling $\sca$ are regarded as nuisance parameters, that should be integrated out.

Finally, we assume that the set of templates $\Fset$ is the linear subspace spanned by the basis vectors $\{\phi_{\ell}\}_{1\leq \ell\leq \dimF}$. Hence, a template $f_{\balpha}\in\Fset$ may be expressed as:
\begin{equation}
\label{eq:template}
f_{\balpha}=\sum_{\ell=1}^{\dimF}\alpha_{\ell}\phi_{\ell}\eqsp,\quad \text{where $\balpha=(\alpha_1,\ldots,\alpha_{\dimF})^{T}\in\alphaSet$,}
 \end{equation}
where for all $\ell\in\{1,\ldots,\dimF\}$, $\phi_\ell:\Uset\to\rset$ and $\alphaSet$ is a subset of $\rset^{m}$.
The pattern is observed at some design points denoted $\design=\{u_1,\ldots,u_{\dimY}\}$, where $\dimY$ is the dimension of the observations such that for all $s\in\{1,\ldots,\dimY\}$, $u_s\in\Uset$. Let $\Phi_{\defo}$ be the $\dimY\times\dimF$ matrix defined such that for all $(s,\ell)$ in $\{1,\ldots,\dimY\}\times\{1,\ldots,\dimF\}$,
\begin{equation}
\label{eq:matPhi}
[\Phi_{\defo}]_{s,\ell}=\phi_{\ell}\circ D(u_s,\defo)\eqsp.
\end{equation}
Defining $\bY=(Y(u_1),\ldots,Y(u_{\dimY}))^{T}$ and $\bW=(W(u_1),\ldots,W(u_{\dimY}))^{T}$ and using \eqref{eq:modelFunc}, the vector of observations can be expressed in a matrix-vector form as:
\begin{equation}
\label{eq:modelVect}
\bY=\sca\Phi_{\defo}\balpha+\sigma\bW\eqsp.
\end{equation}

\subsection{A mixture of deformable templates}
We extend the model to include multiple templates corresponding to the different "typical" shapes that we are willing to cluster and then recognize. To that purpose, we construct a mixture of the template model introduced in the previous section. Denote by $C$ the number of classes $(\cluster{1},\ldots,\cluster{C})$. We associate to each observation $\bY$ an (hidden) class index $I \in \Iset$, where $\Iset=\{1,\dots,C\}$. To each  class $\{\cluster{j}\}_{j\in\Iset}$ is attached a template function $\{f_{j}\}_{j\in\Iset}$ in $\Fset$, which is parameterized by $\{\balpha_{j}\}_{j\in\Iset}\in\rset^{\dimF}$. Moreover, a weight $\omega_j\in(0,1)$ is assigned to the class $I=j\in\Iset$ and we denote by $\bomega=(\omega_1,\ldots,\omega_C)$ the set of prior weights $(\sum_{j=1}^C\omega_j=1)$. To sum up, we consider the following hierarchical model:
\begin{equation}
\label{eq:mixtModelVect}
\bY\in\cluster{j}, \qquad \, \bY=\sca\Phi_{\defo}\balpha_{j}+\sigma\bW\eqsp.
\end{equation}
It is assumed that the observations $\{ \bY_n \}_{n \geq 1}$ are independent random variables, generated as follows:
\begin{equation}
\label{eq:statModel}
\begin{cases}
I_n\sim\multinomial(1,\bomega)\,,\\
\sca_n \sim \gam(a,b)\,,\\
\defo_n\,|\,I_n=j\sim\norm_{\dimDefo}(0_{\dimDefo},\Gamma_j)\eqsp,
\end{cases}
\end{equation}
where $\multinomial$ denotes the multinomial distribution, $(a,b)$ the parameters of the Gamma distribution (assumed known), $0_{\dimDefo}$ the $\dimDefo$-dimensional null vector and $\Gamma_j$ the deformation covariance matrix associated to the $\cluster{j}$. In Section \ref{sec:illustration}, different covariance models are used in function of the deformation model adopted. We stress that the distribution of the scaling parameter is independent of the class index, while the deformation prior distribution is class-dependent. Indeed, on the one hand, the scaling factor accounts for different ranges of observation and is thus independent of what is actually being observed. On the other hand, considering different prior distributions for the deformation might help to learn typical relevant distortions for each class and thus ease the warping process.

In the sequel we assume that $\{ \bW_{n} \}_{n \geq 1}$ is a vector-valued white noise with zero-mean and identity covariance matrix. The extension to more general covariance is straightforward. Hence, conditionally on the class index $I_n$, the global scale $\sca_n$ and local deformation $\defo_n$, the likelihood of $\bY_n$ given the missing data is:
\begin{equation}
\label{eq:obsRV}
  \bY_n\,|\,I_n=j, \sca_n, \defo_n,\sim\,\norm_{\dimY}(\sca_n\Phi_{\defo_n}\,\balpha_{j},\sigma^{2}\mathrm{Id})\eqsp,
\end{equation}
where $\mathrm{Id}$ is the identity matrix.
Denote by $\paramSet$ the set of parameters.
\begin{equation}
\label{eq:paramSP}
\paramSet= \bigcup_{j =1}^{C} \bigg\{\,  \big(\balpha_j,\Gamma_{j},\omega_j,\sigma\big) \, | \, \balpha_j\in\alphaSet,\, \Gamma_{j}\in\posDefMat{},\,\omega_j \in (0,1),\,\sigma>0\,\bigg\} \cap \left\{ \sum_{j=1}^C \omega_j = 1 \right\} \eqsp.
\end{equation}
where $\posDefMat{}$ is the set of $\dimDefo \times \dimDefo$ positive definite matrices.

Let $\bX_n$ be the random vector $\bX_n=(\defo_n,\sca_n)$ taking its values in $\Xset=\defoSpace\times\rset^{+\ast}$ with dimension $\dimX=\dimDefo+1$.  In the sequel, we will use the formalism and the terminology of the  incomplete data model; see \cite{mclachlan:algorithm}. In this formalism, the observation $\bY_n$ stands for the incomplete data, $(I_n,\bX_n)$ are the  missing data and $(I_n,\bX_n,\bY_n)$ are the the complete data. For a given value of the parameter $\param \in \paramSet$, the complete data likelihood $\like_\param$ writes:
\begin{equation}
\label{eq:likelihood}
\like_{\param}(I_n,\bX_n,\bY_n)=g_{\theta}(\bY_n\,|\,I_n,\bX_n)p_{\theta}(\bX_n\,|\, I_n)\omega_{I_n}\eqsp,
\end{equation}
where, for a given value of the parameter $\param\in\paramSet$, $g_\param$ is the conditional density of the observations given the missing data and $p_\param$ is the conditional density of the scaling factor and the local deformation parameter given the class index. Using \eqref{eq:obsRV} and \eqref{eq:statModel}, these densities write
\begin{eqnarray}
\label{eq:prior1}
&&g_{\param}(\bY_n\,|\,I_n,\bX_n)\propto\exp\left(-(1/{2\sigma^{2}})\|\bY_n-\sca_n\Phi_{\defo_n}\balpha_{I_n}\|^{2}\right)\,,\\
\label{eq:prior2}
&&p_{\param}(\bX_n\,|\, I_n)\propto\exp\left(-(1/2) \defo_n^{T} \Gamma_{I_n}^{-1}{\defo_n}\right) \sca_n^{a-1} \exp(-b \sca_n)\eqsp.
\end{eqnarray}
The incomplete data likelihood is obtained by marginalizing the complete data likelihood with respect to the missing data.

\newpage
\section{Sequential parameter estimation using the Online EM algorithm}
\label{sec:onlineEM}
In its original version \cite{Dempster:EM}, the Expectation-Maximization (EM) is a batch algorithm, \ie that uses a fixed set of observations, performing maximum likelihood estimation in incomplete data models. It produces a sequence of parameters, in such a way that the observed likelihood is increased at each iteration. Each iteration is decomposed into two steps. In the E-step, the conditional expectation of the complete data log-likelihood function given the observations and the current fit of the parameters is computed; in the M-step, the parameters are updated by maximizing the conditional expectation computed in the E-step.

In this paper, we focus on a learning setup in which the observations are obtained sequentially and the parameters are updated as soon as a new observation is available. Among several sequential learning algorithms designed to estimate parameters in missing data models, the online EM algorithm proposed in~\cite{Cappe:Online}  sticks closely to the original EM methodology~\cite{Dempster:EM}. It does not require to compute the gradient of the incomplete data likelihood nor the inverse of the complete data Fisher information matrix. Under some mild assumptions, it is shown in \cite{Cappe:Online} that, even when the model is misspecified, the algorithm converges to the set of stationary points of the Kullback-Leibler divergence between the observed likelihood (which does not necessarily belongs to the statistical model) and the incomplete data likelihood. For a given value of the parameter $\param \in \paramSet$, we denote by $\target{\param}{}{\bY_n}$ the posterior distribution of the missing data $(I_n,\bX_n)$, given the observation $\bY_n$. The online EM \cite{Cappe:Online} is initiated with an initial guess $\estim{0}\in\paramSet$. At the $n$-th iteration, the E-step consists in computing the function $\hat{Q}_{n}:\paramSet\to\rset$ defined recursively for all $n>0$ by:
\begin{equation}
\label{eq:estep}
\hat{Q}_{n}(\param)=\hat{Q}_{n-1}(\param)+\varrho_{n}\left(\esp_{\estim{n-1}}\left[\,\log L_{\param}(I_n,\bX_n,\bY_{n})\,|\,\bY_{n}\,\right]-\hat{Q}_{n-1}(\param)\right)\eqsp,
\end{equation}
where $\esp_{\estim{n-1}}(\,\cdot\,|\,\bY_n)$ stands for the conditional expectation under $\target{\estim{n-1}}{}{\bY_n}$, $\{\varrho_{n}\}_{n>0}$ is a decreasing sequence of positive step sizes, with $\varrho_1=1$, such that $\hat{Q}_0$ needs not be specified. In the M-step, the next estimate $\estim{n}$ is obtained by maximizing
\begin{equation}
\label{eq:mstep}
\estim{n}=\mathrm{arg}\,\max\limits_{\param \in \paramSet}\, \hat{Q}_{n}(\param)\eqsp.
\end{equation}

Under our model specification, the complete data log-likelihood belongs to a curved exponential family. Indeed, for a given parameter $\param\in\paramSet$, $\log\like_\param$ writes
\begin{equation}
\label{eq:exp_model}
\log\like_\param(I,\bX,\bY)=t(\param)+\pscal{}{r(\param)}{S(I,\bX,\bY)}\eqsp,
\end{equation}
where the function $t$ is given by
\begin{equation*}
\label{eq:def_t}
t(\param)=\log\frac{b^a}{\gammaFct(a)}-\frac{\dimY}{2}\log{2\pi\sigma^2}-\dimDefo\log{2\pi}\,,
\end{equation*}
and the functions $r(\param)=(r_1(\param),\ldots,r_C(\param))$ and $S(I,\bX,\bY)=(S_1(I,\bX,\bY),\ldots,S_C(I,\bX,\bY))$, such that for all $j\in\{1,\ldots,C\}$:

\begin{gather*}
r_j(\param)=(1/2)\left(
  2\log(\omega_j)-\log\det\Gamma_j,
  2\sigma^{-2}\balpha_j,
  -\sigma^{-2}(\balpha_j\balpha_j^{T}),
  -{\Gamma_j^{-1}}^{T},
  -\sigma^{-2},
  -2b,
  2(a-1)\right)\,,\\
S_j(I,\bX,\bY) =\delta_{I,j}\left(
  1,
  \sca\phi_{\defo}^{T}\bY,
 \sca^2\phi_{\defo}^{T}\phi_{\defo},
  \defo\defo^{T},
  \|\bY\|^{2},
  \sca,
  \log{\sca}
  \right)\,.
\end{gather*}

As a consequence, the two steps of the online EM consist in (i) computing for all $j\in\{1,\ldots,C\}$ the stochastic approximation (SA) recursion
\begin{align}
\label{eq:estep_curvedExp}
&\hat{s}_{n,j}=\hat{s}_{n-1,j}+\varrho_{n}\left(\bar{s}_{n,j}(\bY_{n};\estim{n-1})-\hat{s}_{n-1,j}\right),
\end{align}
where $\bar{s}_{n,j}(\bY_n;\estim{n-1})= \esp_{\estim{n-1}}\left[\,S_j(I_n,\bX_n,\bY_{n})\,|\,\bY_n\right]$ and (ii) updating the parameters according to
\begin{align}
\label{eq:mstep_curvedExp}
&\estim{n}=\mathrm{arg}\,\max\limits_{\param \in \paramSet}\;\left\{t(\param)+\sum_{j=1}^{C}\pscal{}{r_j(\param)}{\hat{s}_{n,j}}\right\}\,.
\end{align}
The maximization is in closed form. However, this algorithm remains essentially of theoretical interest because in our model the conditional expectation $\bar{s}_{n,j}(\bY_n;\estim{n-1})$ is not analytically tractable. Intractable E-steps have already been addressed for batch EM algorithms. In \cite{Delyon:ConvSAEM}, the authors proved the convergence of the Stochastic Approximation EM (SAEM) algorithm in which the E-step is replaced by a stochastic approximation making use of realizations of the missing data generated according to the posterior distribution. Still, extending the SAEM algorithm to the online setup is not feasible in our case. Indeed, independent and identically distributed (\iid) samples from $\target{\estim{n-1}}{}{\bY_n}$ can not be simulated. An alternative to the SAEM algorithm, known as MCMC-SAEM, was proposed in \cite{Kuhn:MCMCSAEM}: the authors suggested to use Markov chain Monte Carlo (MCMC) methods (see \cite{Andrieu:IntroMCMC} for an introduction) to obtain samples from the posterior distribution.

In this paper, we adapt this approach to the sequential setting outlined above leading to the MCoEM (Monte Carlo online EM) algorithm. It is a 3-step iterative algorithm. Given the current fit of parameter $\estim{n-1}$ and a new observation $\bY_{n}$, the algorithm proceeds as follows:
\begin{enumerate}[(1)]
  \item \emph{simulation step}: simulate, using a $\target{\estim{n-1}}{}{\bY_n}$-reversible Markov kernel $\K{n}$, a Markov chain $\{I_n[k],\bX_{n}[k]\}_{k>0}$,
  \item \emph{stochastic approximation step}: update for each class $j\in\{1,\ldots,C\}$, the complete data sufficient statistics using the following recursion
  \begin{equation}
  \label{eq:estep_SAOEM} \tilde{s}_{n,j}=\tilde{s}_{n-1,j}+\varrho_{n}\left(\frac{1}{m_n}\sum_{k=1}^{m_n}S_j(I_n[k],\bX_n[k],\bY_n)-\tilde{s}_{n-1,j}\right),
    \end{equation}
  where $m_n$ is the number of MCMC iterations performed at the $n$-th iteration of the MCoEM algorithm,
  \item \emph{maximization step}: update the parameter $\estim{n}$ by maximizing the function~:
  \begin{equation}
  \label{eq:mstep_SAOEM}
  \estim{n}=\mathrm{arg}\,\max\limits_{\param \in \paramSet}\;\left\{t(\param)+ \sum_{j=1}^{C}\pscal{}{r_j(\param)}{\tilde{s}_{n,j}}\right\}\,.
  \end{equation}
\end{enumerate}

For numerical stability, it is recommended not to update the parameter $\estim{n}$ at each iteration, especially in the first iterations of the algorithm (see discussion in Section \ref{sec:illustration}). MCoEM updates $\estim{n}$ according to an user-defined update schedule $\mathfrak{N}\subset \nset$. Algorithm \ref{alg:MCoEM} provides a pseudo-code representation of MCoEM.

\begin{algorithm}
\caption{Monte Carlo online EM}
\label{alg:MCoEM}
\begin{algorithmic}[1]
 \State {\bf{Input:}}
        \begin{itemize}
			\item Initial guess: $\estim{0}\in\paramSet$
            \item A stream of observations: $\bY_1,\bY_2,\ldots$				
            \item Parameter update schedule: $\mathfrak{N}\subseteq \nset$
            \item An iteration counter $n$, initialized to $0$
            \item A sequence of positive step sizes $\{\varrho_1,\varrho_2,\ldots\}$ with $\varrho_1=1$
            \item MCMC length schedule $\{m_1,m_2,\ldots\}$
		\end{itemize}	

\vspace{.5cm}

\State {\textbf{When} a new observation $\bY$ is available \textbf{do}}
    \State \hspace{.7cm} Increment the iteration counter: $n=n+1$
    \State \hspace{.7cm} \textbf{Simulation step}: Sample $m_n$ missing data $\{I_{n}[k],\bX_{n}[k]\}_{k=1}^{m_n}$ from a Markov chain targeting $\pi_{\estim{n-1}(\,\cdot\,|\,\bY)}$

    $\blacktriangleright$ See Algorithm \ref{alg:MCMC}
    \State \hspace{.7cm} \textbf{SA step}: Update the sufficient statistics $\tilde{s}_{n,1},\ldots, \tilde{s}_{n,C}$ via the stochastic approximation step

    $\blacktriangleright$ See Eq. \eqref{eq:estep_SAOEM}
    \State \hspace{.7cm} {\textbf{If} $n\in\mathfrak{N}$ \textbf{then}}
    \State \hspace{1.4cm} \textbf{Maximization step}: Update the parameter estimate to $\estim{n}$

    \hspace{.7cm}$\blacktriangleright$ See Eq. \eqref{eq:mstep_SAOEM}
    \State \hspace{1.4cm}{\textbf{else}}
    \State \hspace{1.4cm}Set $\estim{n}=\estim{n-1}$
	\State \hspace{.7cm}{\textbf{end if}}

\vspace{.5cm}
\State {\bf{Output:}} A sequence of parameters $\estim{1},\estim{2},\ldots$.

\end{algorithmic}
\end{algorithm}

\section{Approximating the hidden data joint posterior distribution}
\label{sec:carlinChib}
In this section, we construct a transition kernel $\K{}$ to sample the target distribution $\target{\param}{}{\bY}$ (for notational simplicity, the iteration index $n$ of the EM algorithm is omitted in this section).

\begin{rem}
At this stage, one might legitimately wonder why a special care must be taken when choosing $\K{}$, while valid MCMC routines are by now well established and available. Having a closer look at the target distribution dismisses resorting to standard MCMC methods such as the Gibbs sampler \cite{Gelfand:GibbsSampling,Geman:Gibbs} to simulate samples from $\target{\param}{}{\bY}$. Indeed, the target distribution is not defined on the product space $(\Iset,\Xset)$ but on the following union of spaces $(\Iset=1,\Xset)\cup\cdots\cup (\Iset=C,\Xset)$. This is because, in our framework, the deformation $\bX$ should always be consistent with the class of the observation it applies to.
\end{rem}

\subsection{MCMC on an extended state space}
\label{subsec:mcmc}
We now explain the approach we followed. The basic idea, stemming from \cite{Carlin:Bayesian}, is to specify a joint distribution over the class index $I$ and auxiliary variables $\bU_{1},\ldots, \bU_{C}$, where for all $j\in\{1,\ldots,C\}$, $\bU_{j}\in\Xset$ is a deformation parameter associated to the class $\cluster{j}$. We stress that, in this approach, we sample at each iteration deformation parameters for each class. To specify the joint distribution, we introduce the \textit{pseudo-priors} or \textit{linking densities}, denoted $\{\kappa_{\param,j}\}_{j=1}^{C}$. Note that whereas the knowledge of the normalizing constant is not required for an MCMC algorithm, the normalizing constant of the pseudo-priors are assumed to be known, \ie\ the pseudo-priors $\{\kappa_{\param,j}\}_{j=1}^{C}$ should integrate to $1$. Also, it is assumed that exact sampling from the pseudo-priors is doable (and is computationally inexpensive). We define an auxiliary joint posterior density $\targetCC{\param}{}{\bY}$ on the product space $\Iset\times\Xset\times\cdots\times\Xset$ by:
\begin{multline}
\targetCC{\param}{I,\bU_1,\ldots,\bU_C}{\bY}=\target{\param}{I,\bU_I}{\bY}\prod_{j\neq I}\kappa_{\param,j}(\bU_j)\\
\propto g_{\param}(\bY\,|\,I,\bU_I)p_{\param}(\bU_I\,|\,I)\omega_I\prod_{j\neq I}\kappa_{\param,j}(\bU_j)\eqsp,\label{eq:targetCC}
\end{multline}
where $\omega_I$, $g_{\param}$ and $p_{\param}$ are defined  in~\eqref{eq:statModel},~\eqref{eq:prior1} and \eqref{eq:prior2} respectively. It can be noted that the marginal of $\targetCC{\param}{}{\bY}$ \wrt\ to the auxiliary deformation parameters is the target distribution $\target{\param}{}{\bY}$:
\begin{equation}
\label{eq:marginale}
\target{\param}{I,\bX}{\bY}=\idotsint \targetCC{\param}{I,\bu_{1:I-1},\bX,\bu_{I+1:C}}{\bY}\rmd \bu_{-I}\eqsp,
\end{equation}
where for all $(i,j)\in\Iset^{2}$, such that $i<j$, $a_{i:j}=(a_i, a_{i+1}, \dots, a_j)$ and for all $i\in\Iset$, $a_{-i}=\{a_{j}\}_{j=1,j\neq i}^{C}$. Remarkably, this property does not depend on the choice of pseudo-priors.

A Metropolis-within-Gibbs sampler targeting $\targetCC{\param}{}{\bY}$ is used to simulate a Markov chain  $(I[k],\bU_1[k],\ldots,\bU_C[k])$ on the product space $(\Iset\times\Xset\times\ldots\allowbreak\times\Xset)$. Suppose the Markov chain is at state $(I,\bU_1,\ldots,\bU_C)$, the so-called full conditional posterior distributions required for the Gibbs sampler are:
\begin{align}
\label{eq:discreteCC}
&\targetCC{\param}{I}{\bU_{1:C},\bY}\propto g_{\param}(\bY\,|\,I,\bU_{I})p_{\theta}(\bU_{I}\,|\,{I})\omega_{I}\prod_{j\neq I}\kappa_{\param,j}(\bU_j)\,.\\
\label{eq:auxiliaryCC}
&\targetCC{\param}{\bU_j}{I,\bU_{-j},\bY}\propto
\begin{cases}
g_{\param}(\bY\,|\,I,\bU_{I})p_{\param}(\bU_{I}\,|\,I)=\pi_{\param}(\bU_I\,|\,I,\bY)\eqsp,&j=I\\
\kappa_{\param,j}(\bU_j)\eqsp,&j\neq I\,.
\end{cases}
\end{align}

From \eqref{eq:discreteCC} and \eqref{eq:auxiliaryCC}, it can be seen that sampling the class index and the auxiliary deformations from their respective full conditional posterior distribution is straightforward. However, since sampling the new parameter from the current class cannot be achieved directly, a Random Walk Metropolis-Hastings (RWMH) \cite{Metropolis:metropolisAlg} kernel $P_{\param}(\bU_I;\,\cdot\,|\,\bU_{-I},I,\bY)$ having $\pi_\param(\,\cdot\,|\,I,\bY)$ as its stationary distribution is applied $r$ times to $\bU_I$ to generate $\bU'_I$. The Markov chain transition writes:
\begin{enumerate}[(i)]
\item $I'\sim \targetCC{\param}{\cdot}{\bU_{1:C},\bY}$
\item $\bU_j'\sim \kappa_{\param,j}$, for $j\neq I'$
\item $\bU_{I'}'\sim P^r_{\param}(\bU_{I'};\,\cdot\,|\,\bU_{-I'}',I',\bY)$
\end{enumerate}
and the transition kernel $\tK{}^{\CC}$ may thus be expressed as:
\begin{equation}
\label{eq:kernelCC_extended}
\tK{}^{\CC}(I',\rmd\bU_{1:C}'\,|\,I,\bU_{1:C})=\targetCC{\param}{I'}{\bU_{1:C},\bY}P_{\param}(\bU_{I'};\rmd \bU_{I'}'\,|\,\bU_{-I'}',I',\bY)\prod_{j\neq I}\kappa_{\param,j}(\rmd\bU_j')\eqsp.
\end{equation}

The Markov chain $\{I[k],\bU_1[k],\ldots,\bU_C[k]\}_{k>0}$, simulated through a Metropolis-within-Gibbs algorithm, provides samples from $\targetCC{\param}{}{\bY}$. However, only the marginal samples $\{I[k],\bX[k]=\bU_{I[k]}\}_{k>0}$, distributed under $\target{\param}{}{\bY}$ \eqref{eq:marginale}, are of interest and will be used in the approximation of the E-step of the MCoEM algorithm \eqref{eq:estep_SAOEM}. Pseudo-code of the Markov chain simulation algorithm is reported in Algorithm \ref{alg:MCMC}.

\begin{algorithm}
\caption{Markov chain simulating missing data}
\label{alg:MCMC}
\begin{algorithmic}[1]

 \State {\bf{Input:}}
        \begin{itemize}
			\item An observation: $\bY$
            \item A parameter estimate: $\param$	
            \item Number of components: $C$
            \item Length of the Markov chain: $m$
            \item Number of RWMH iterations: $r$
		\end{itemize}	

\vspace{.5cm}
\State Specification of the pseudo-prior densities $\kappa_{\param,1},\ldots,\kappa_{\param,2}$

 $\blacktriangleright$ See Section \ref{subsec:pseudo}

\State Set $\bU_j[0]\sim \kappa_{\param,j}$ for $j=1,\ldots,C$

\For{$k=1,\ldots,m$}
\State Class sampling: $I[k]\sim \targetCC{\param}{I}{\bU_{1:C}[k-1],\bY}$

 $\blacktriangleright$ See Eq. \ref{eq:discreteCC}

\State Let $i=I[k]$
\State Random Walk Metropolis-Hastings move: $\bU_{i}[k]\sim P^r_{\param}(\bU_{i}[k-1];\,\cdot\,|\,i,\bY)$
\For {$j\in\{1,\ldots,C\}\backslash\{i\}$}
\State Pseudo-prior update: $\bU_j[k]\sim\kappa_{\param,j}$
\EndFor
\State Set $\bX[k]=\bU_{i}[k]$
\EndFor

\State {\bf{Output:}} A Markov chain $(I[1],\bX[1],\ldots,I[m],\bX[m])$.

\end{algorithmic}
\end{algorithm}

\subsection{Choice of the pseudo-prior densities}
\label{subsec:pseudo}
The specification of the linking densities is essential for sampling efficiency. Ideally, these densities should be close to the marginal posterior: for all $j\in\{1,\ldots,C\}$, the density $\bX \to \kappa_{\param,j}(\,\bX\,)$ should be chosen as a proxy to $\bX\to\pi_{\theta}(\bX\,|\,j,\bY)$. An idea is for instance to set the pseudo-prior density as a Gaussian approximation of the target density. Such an approximation can be obtained using the Laplace method~\cite{Wolfinger:Laplace} or other approximate Bayesian sampling method. Under the (weak) assumption that the function $\bX\to\target{\theta}{\bX}{j,\bY}$ admits a maximum,
\begin{equation}
\label{eq:pseudo_mean}
\bXs_j=\arg\max_{\bX\in\Xset}\pi_{\theta}(\bX\,|\,j,\bY)\,,
\end{equation}
the Taylor-expansion of the logarithm of $\pi_{\theta}(\bX\,|\,j,\bY)$ writes:
\begin{equation}
\label{eq:TaylorExp}
\log{\pi_{\theta}(\bX\,|\,j,\bY)}=\log{\pi_{\theta}(\bXs_j\,|\,j,\bY)}+\frac{1}{2}(\bX-\bXs_j)^{T}H_{j} (\bX-\bXs_j)+o(\|\bX-\bXs_j\|^{2})\,,
\end{equation}
where for all $j\in\{1,\ldots,C\}$, $H_j$ is the Hessian matrix, whose coefficients are given for all $(q,r)\in\{1,\ldots,\dimX\}^{2}$ by:
\begin{equation}
\label{eq:pseudo_cov}
[H_j]_{q,r}=\left.\frac{\partial^{2}}{\partial \bX_{q}\partial \bX_{r}}\log{\pi_{\theta}(\bX\,|\,j,\bY)}\right|_{\bX=\bXs_j}\,.
\end{equation}
Note that for better readability, for all $j\in\{1,\ldots,C\}$, the dependence of the linking densities $\kappa_{\param,j}$, and the parameters $\bXs_j$, $H_j$ on $\bY$ and $\param$ is not made explicit in these notations, but does exist.

The previous discussion suggests that $\norm_{\dimX}(\bXs_j,-H_{j}^{-1})$ is a sensible candidate for $\kappa_{\param,j}$. The pseudo-priors parameters $\bXs_j$ may be obtained using standard nonlinear optimization methods. Since $\bXs$ is only used in the pseudo-prior specification, the precision of the optimizer does not matter much and simple heuristics can be used (see related discussion in Section \ref{sec:illustration}).

\begin{rem}
Our proposed kernel shares some similarities with that proposed in \cite{Allassonniere:Mixture}, which also makes use of auxiliary variable $\{\bU_1,\ldots,\bU_C\}$. These authors propose to first sample the class index $I$ from $\pi_{\param}(\,\cdot\,|\,\bY)$ and then draw $\bX\sim\pi_{\param}(\,\cdot\,|\,I,\bY)$. However, since sampling the class index from the posterior distribution is not doable (indeed $\pi(I=j\,|\,\bY)\propto\pi_\param(j,\bY)$ which is not analytically tractable), auxiliary variables $\{\bU_1[k],\ldots,\bU_C[k]\}_{k>0}$ are sampled from $C$ independent Markov chains each targeting $\pi_{\param}(\,\cdot\,|\,j,\bY)$, $j\in\{1,\ldots,C\}$, in an attempt to approximate the posterior weights $\{\pi_{\param}(j\,|\,\bY)\}_{j=1}^{C}$. These approximate weights allow to sample $I$ and then parameter samples $\{\bX[k]\}_{k>0}$ are drawn using a Markov chain targeting $\pi_\param(\,\cdot\,|\,I,\bY)$. This scheme is computationally intensive all the more that \cite{Allassonniere:Mixture} uses a batch learning setup which implies that at each iteration, as many latent variables $\{I_n[k],\bX_n[k]\}_{k>0}$ as there are observations in the dataset need to be sampled.
\end{rem}

\section{Numerical Illustration}
\label{sec:illustration}
We evaluate the performance of our online learning algorithm by inferring two types of data: growth velocity curves and handwritten digits. These two examples illustrate the flexibility, the stability and the computational effectiveness of the proposed MCoEM. MCoEM is then compared to an equivalent SAEM algorithm on the handwritten digits templates extraction task.

\subsection{Growth velocity curve study}
\label{subsec:5:1}
\begin{figure}
    \centering
    \includegraphics[scale=1]{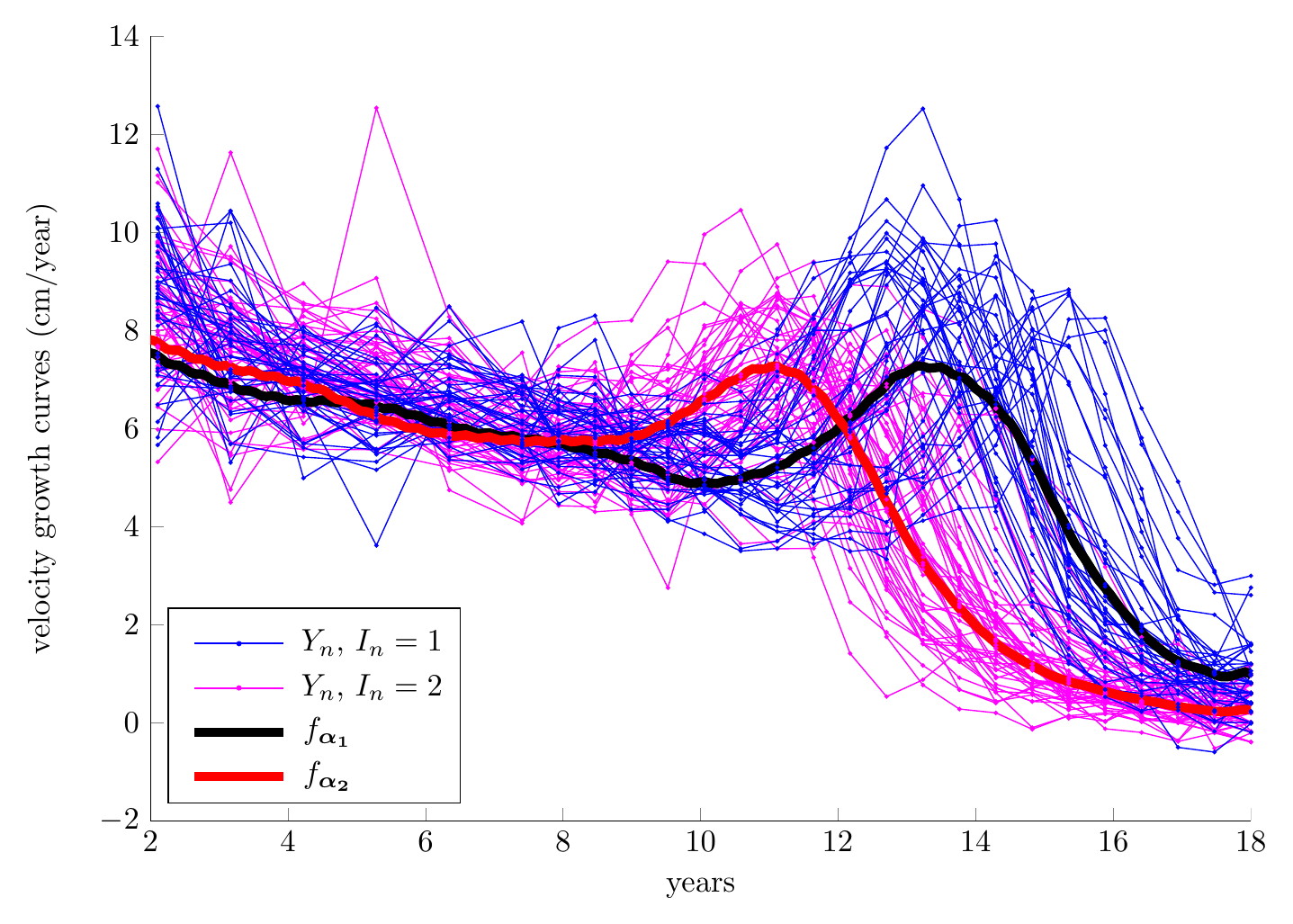}
    \caption{Growth velocity samples and templates extraction obtained through $1,000$ iterations of the MCoEM algorithm.\label{fig:curves}}
\end{figure}

The growth velocity curve example is a classical benchmark in curve registration; see \cite{ramsay:functional},\cite{zhong:2008}. It is used here for illustrative purposes, because  the rationale of the model is easy to grasp. The growth curves are obtained from the Berkeley Growth Study data \cite{tuddenham:physical} and display the evolution of the growth velocity between $2$ and $18$ years, for $39$ boys and $54$ girls; see Figure \ref{fig:curves}. Even though each observation is known to arise from either a boy or a girl, we won't make any use of this information, as MCoEM is designed for unsupervised inference on mixture models. The objective of the algorithm is therefore to retrieve a standard growth profile for boys and girls from the unlabeled set of growth velocity curves. The growth velocity curves, plot the growth velocity of individuals observed at $\dimY=31$ landmarks $\design=\{u_{1},\ldots,u_{\dimY}\}$, irregularly spaced, such that for all $s\in\{1,\ldots,\dimY\}$, $2\leq u_s\leq 18$.

\subsubsection{Deformable template model}
Growth profiles may vary from an individual to another, both as a function of the time and in amplitude. The algorithm aims to extract templates for the growth velocity curves: it associates to each observation $Y_n$ a monotonically increasing time warping function $u\mapsto D(u,\defo_n)$ as well as a global scaling parameter $\sca_n$. We consider a mixture model with $C=2$, implying that we aim at retrieving templates for boys and girls growth velocity separately: the class index $I_n\in\{1,2\}$ models the boys and girls clusters. In this illustration, the template is a function $f_{\balpha_i}$ ($i\in\{1,2\}$) defined on an open segment $\Uset=\ooint{u_i}{u_f}=\ooint{2}{18}$ parameterized as:
\begin{equation}
\label{eq:curveRegTemplate}
f_{\balpha_i}(u)=\sum_{\ell=1}^{m}\alpha_{i,\ell}\phi_\ell(u)\eqsp,\quad (\alpha_{i,1},\ldots,\alpha_{i,\dimF})\in\alphaSet={\rset^{+}}^{\dimF}\,,
\end{equation}
where $\{\phi_{\ell}\}_{\ell=1}^{\dimF}$ is set as $u\mapsto\phi_\ell(u)=\exp{({\nu}_\ell^{-2}(u-r_\ell)^2)}$, where $\{r_\ell\}_{\ell=1}^{\dimF}$ are regularly spaced landmark points in $\Uset$. The choice of $\{\phi_\ell\}_{\ell=1}^{\dimF}$ and $\alphaSet$ ensures that the template function $u\mapsto f_{\balpha_i}(u)$ is a positive function, which is a natural constraint for growth velocity curves. For all $\ell\in\{1,\ldots,\dimF\}$, the bandwidth of $\phi_\ell$ is set as ${\nu}_\ell^2=-\frac{\min_{u\in\design}\|r_\ell-u\|^{2}}{\log{\varepsilon}}$, where $\varepsilon \in \ooint{0}{1}$ is the value of $\phi_\ell$ at the nearest design point of $r_\ell$. This choice of bandwidth enables to take into account the irregularly spaced measurement points in $\design$. In this implementation, we used $m=35$, so that kernels $\phi_1,\phi_2,\ldots$ are centered on landmarks distant from a 6-month interval and $\varepsilon=0.1$. The deformable template model \eqref{eq:modelFunc} simply writes for all $u\in\Uset$:
$$
Y_n\in\cluster{i},\qquad Y_n(u)=\lambda_n f_{\boldsymbol{\alpha_i}}\circ D(u,\defo_n)+\sigma W_n(u)\,,
$$

In this setting, the time warping function $u\mapsto D(u,\defo)$ is monotonically increasing and should satisfy $D(u_i,\defo)\geq u_i$ and $D(u_f,\defo)\leq u_f$ (indeed, outside $\ooint{u_i}{u_f}$, the template vanishes \eqref{eq:curveRegTemplate}). In order to satisfy these constraints, we write $D(\,\cdot\,,\defo)$ as:
\begin{equation}
\label{eq:locDef_curves}
D(u,\defo)=u_i+(u_f-u_i)H(u,\defo)\eqsp,
\end{equation}
where $H(\,\cdot\,,\defo)$ is modeled as proposed in \cite{Ramsay:curveReg} with:
\begin{equation}
\label{eq:locDef_H}
H(u,\defo)=\frac{\int_{{u'_i}}^{u}\exp{\left[\sum_{k=1}^{\dimDefo}\defo_k\psi_k(v)\right]\rmd v}}{\int_{u'_i}^{u'_f}\exp{\left[\sum_{k=1}^{\dimDefo}\defo_k\psi_k(v)\right]\rmd v}}\eqsp,
\end{equation}
where $u'_i\leq u_i$ and $u'_f\geq u_f$ allow to satisfy the constraints stated above. For all $k$ in $\{1,\ldots,\dimDefo\}$, $\defo_k\in\rset$ and $\{\psi_k\}_{k=1}^{\dimDefo}$ is a dictionary of Gaussian kernels centered on the landmark points $\{q_k\}_{k=1}^{\dimDefo}$ with the same bandwidth $\tau^2$.  In this implementation, we set $u'_i=0$, $u'_f=20$ and use $\dimDefo=20$ regularly spaced landmark points such that $q_1=u'_i$ and $q_{\dimDefo}=u'_f$; the kernel variance is set to $\tau^{2}=1$. Moreover, the prior distribution \eqref{eq:prior1} of $\defo$ is set with a mean equals to $(1,\dots,1)^T$ and for all $j\in\{1,2\}$ a covariance matrix $\Gamma_j$ parameterized by the variance $\gamma_j$, such that $\Gamma_j=\gamma_j^{2} \mathrm{Id}_{\dimDefo}$. The estimate $\hat{\gamma}^{2}_{j,n}$ of $\gamma_j^{2}$ after $n=1000$ iterations is $\hat{\gamma}_{1,1000}^{2}=0.08$ and $\hat{\gamma}_{2,1000}^{2}=0.07$. A Gamma prior with parameters $a=b=10$ is assumed for $\sca_n$.

\subsubsection{Sampling the missing data}
\begin{figure}
\centering
\begin{tabular}{c}
\includegraphics[width=.8\textwidth]{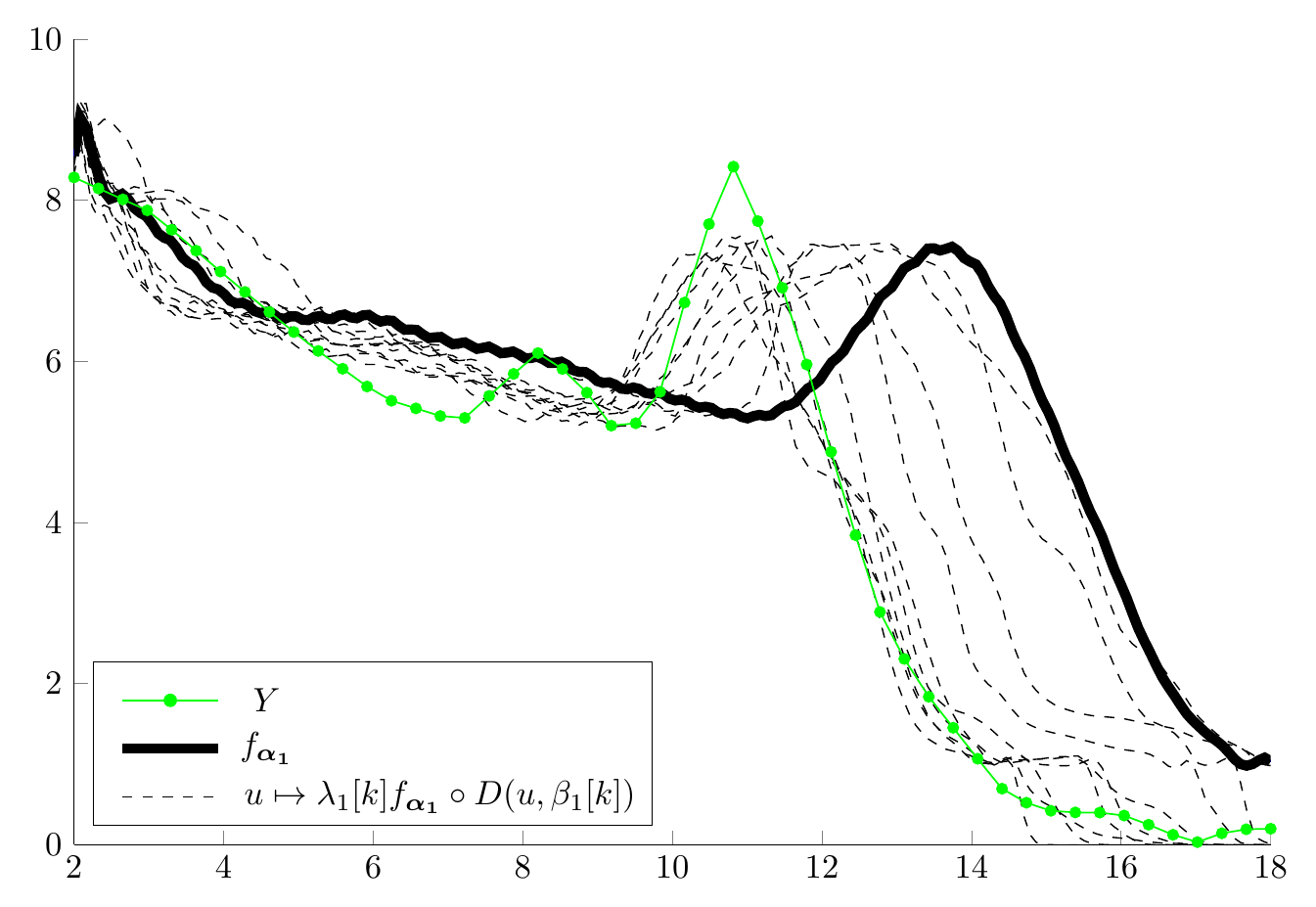} \\
\includegraphics[width=.8\textwidth]{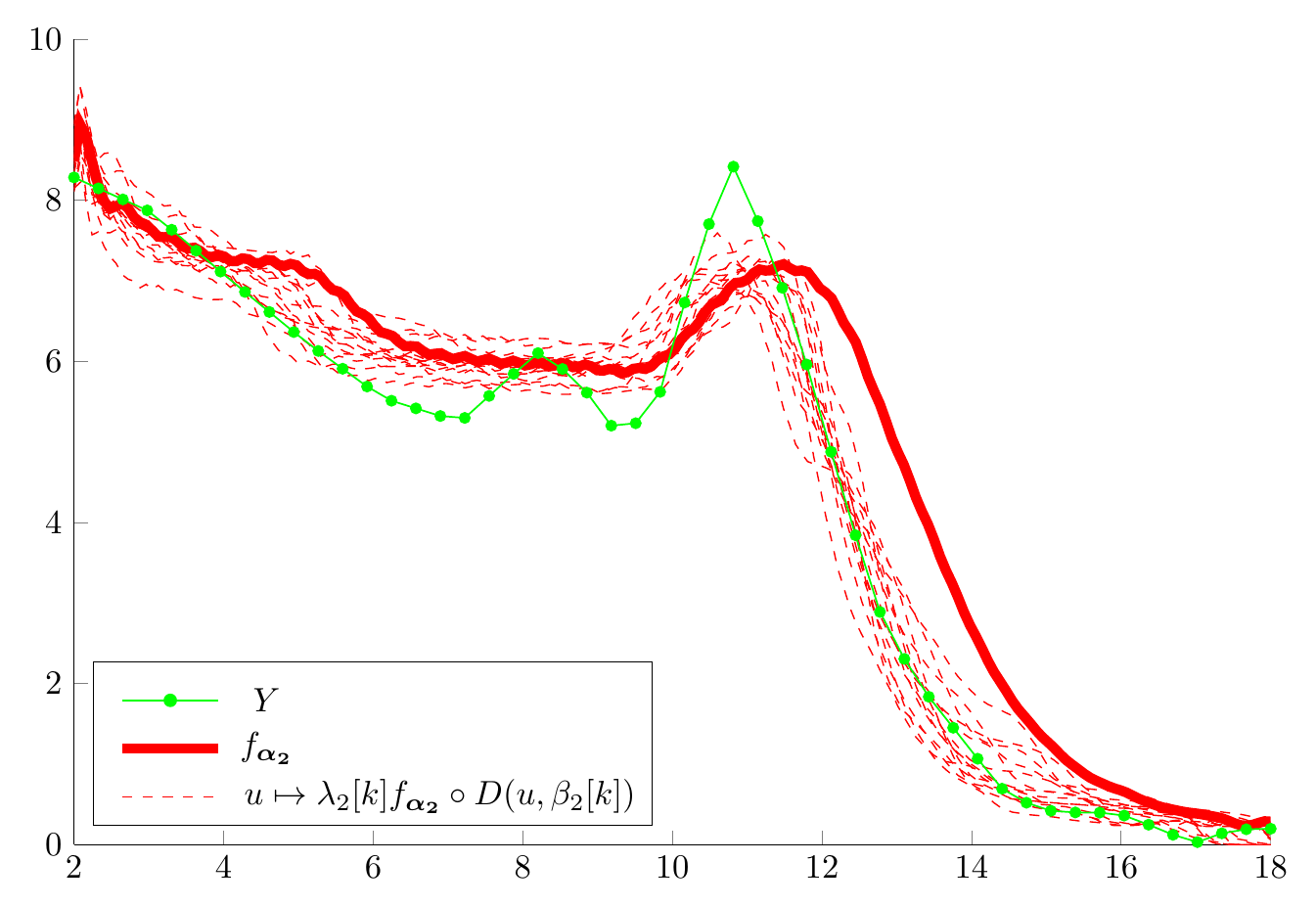}
\end{tabular}
\caption{Sampling of the hidden data posterior distribution. $\{\lambda_1[k],\beta_1[k],\lambda_2[k],\beta_2[k],I[k]\}$ for some $k\in\{101\ldots,300\}$ are samples from the Markov chain produced by $\tK{}^{\CC}$ that admits $\tilde{\pi}_{\param}(\,\cdot\,|\,\bY)$ \eqref{eq:targetCC} as stationary distribution. The sampled deformation/scale are then applied to the template of the class it corresponds to (the thick black/red line), yielding a distorted template (the dashed black/red line) that tends to match the observation (the green line). \label{fig:curveSamplingScheme_1}}
\end{figure}

\begin{figure}
\centering
\includegraphics[scale=.9]{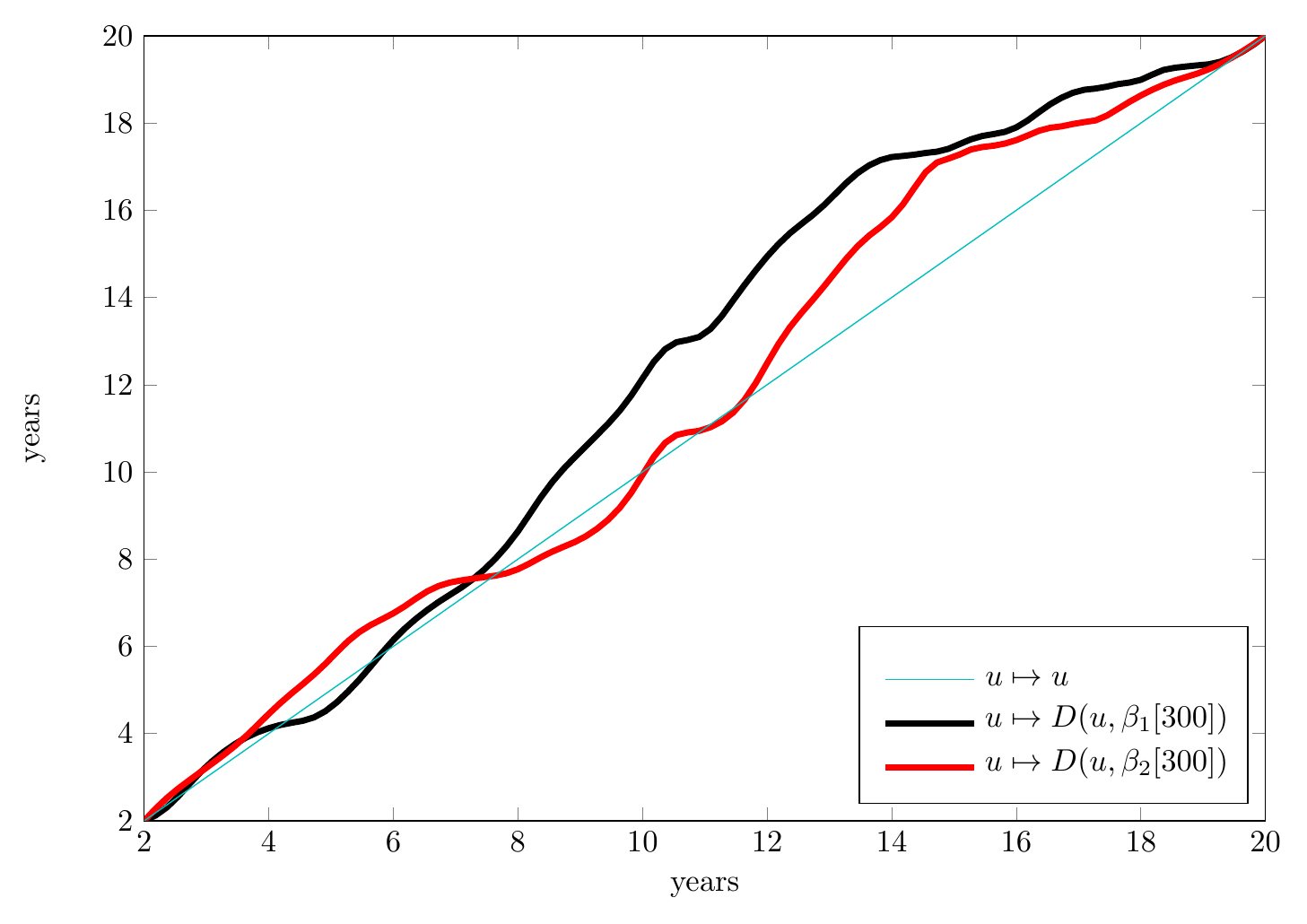}
\caption{Time warping functions for the deformation parameters $\beta_1$ and $\beta_2$ sampled at the last iteration ($k=300$) of the Markov chain produced by $\tK{}^{\CC}$ that admits $\tilde{\pi}_{\param}(\,\cdot\,|\,\bY)$ \eqref{eq:targetCC} as stationary distribution; see Figure \ref{fig:curveSamplingScheme_1}. \label{fig:curveSamplingScheme_2}}
\end{figure}

\begin{figure}
\centering
\includegraphics[scale=.85]{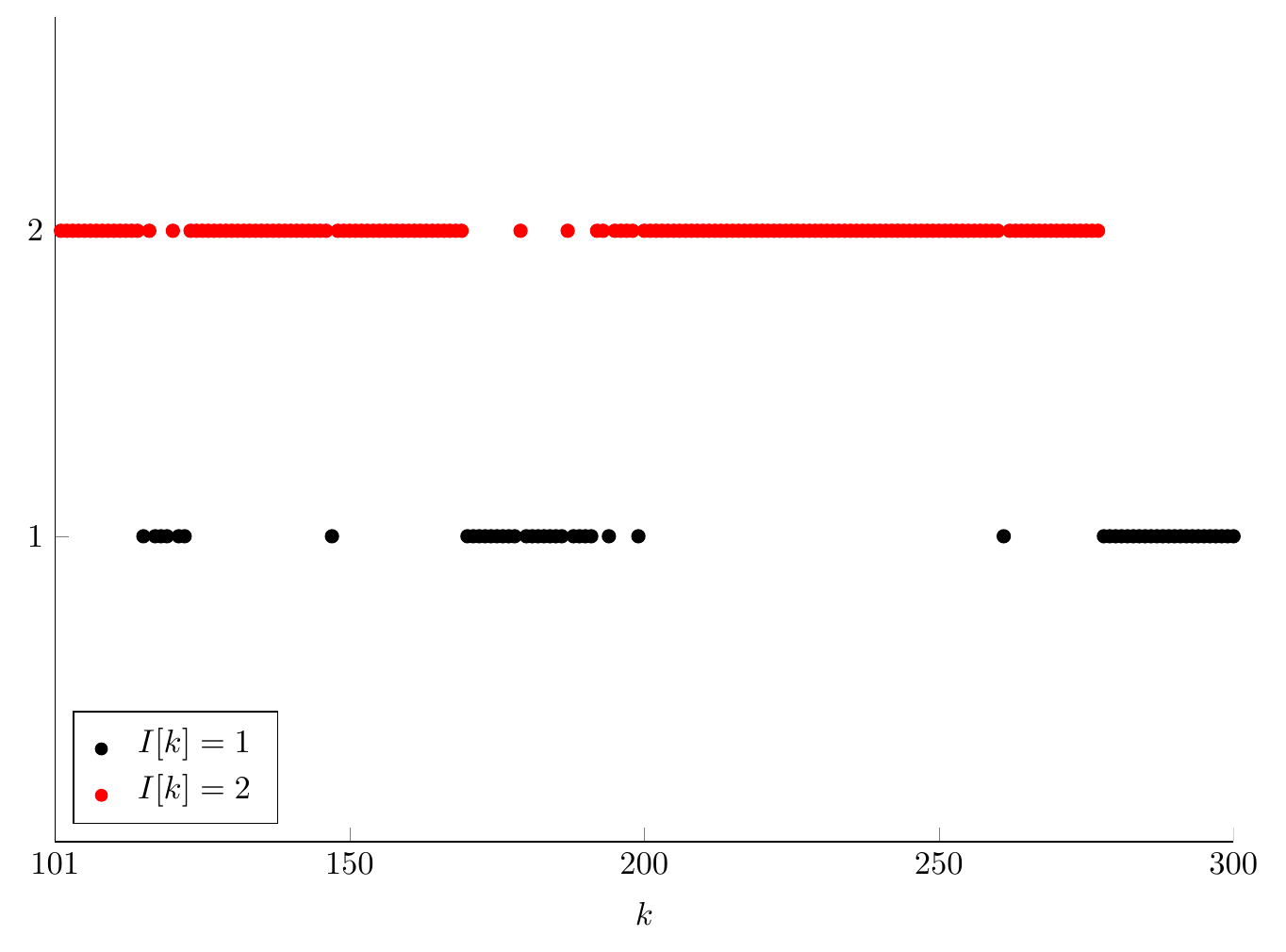}
\caption{Class sampling $\{I[1],\ldots,I[300]\}$ from the Markov chain produced by $\tK{}^{\CC}$ that admits $\tilde{\pi}_{\param}(\,\cdot\,|\,\bY)$ \eqref{eq:targetCC} as stationary distribution. \label{fig:curveSamplingScheme_3}}
\end{figure}

The Figures \ref{fig:curveSamplingScheme_1}--\ref{fig:curveSamplingScheme_3} illustrate the sampling scheme proposed in Section \ref{sec:carlinChib}, taking place at a given iteration $n$ of the MCoEM algorithm (the index $n$ is omitted hereafter). For $j\in\{1,2\}$, the auxiliary variable $\bU_j$ consists in $\bU_j=(\sca_j,\defo_j)$. In Figure \ref{fig:curveSamplingScheme_1}, green dots represent an observation $Y$ along with the templates in plain curves (boys on the top panel and girls on the bottom panel). In each panel, the dashed curves illustrate different realizations of the distorted template under the action of deformation parameters $\bU_1[k]=(\defo_1[k],\sca_1[k])$ and $\bU_2[k]=(\defo_2[k],\sca_2[k])$ sampled using the kernel $\tK{}^{\CC}$. For each new observation $\bY$, we used $300$ iterations of the Markov chain detailed in Subsection \ref{subsec:mcmc}, discarding the first 100 states for burn-in. The pseudo-priors $\kappa_1$ and $\kappa_2$ were set as Gaussian distributions, as specified in Subsection \ref{subsec:pseudo}. For $j\in\{1,2\}$, the mean $(\sca^{\star}_j,\defo^{\star}_j)$ were obtained through a quasi-Newton optimization method (with an early stopping rule, because the precision of the fit does not matter much). For computational efficiency, the covariance matrix was set as $\hat{\Gamma}_{j,n}=\hat{\gamma}_{j,n}^2\mathrm{Id}_{\dimDefo}$ (which is the $j$-th class prior covariance matrix estimate). Even though, the pseudo-prior distributions provide inappropriate deformation parameters (see some samples from $D(\,\cdot\,,\defo_{1}[k])$ on the top panel of Figure \ref{fig:curveSamplingScheme_1}), they nevertheless achieve their two-fold target, namely (i) allowing to switch between models as illustrated in Figure \ref{fig:curveSamplingScheme_2} and (ii) sampling deformations that are consistent with $\bY$: the distorted templates tend to match the observation. Figure \ref{fig:curveSamplingScheme_2} shows two warping functions $D(\,\cdot\,,\defo_1[k])$  and $D(\,\cdot\,,\defo_2[k])$ corresponding to the samples $\defo_1[k]$ and $\defo_2[k]$ obtained at the $k=300$-th iteration of the Markov chain. This shows that, in order to register the template with the observation, the boys time warping function (in black, parameterized by $\beta_1$) accelerates the time from 9 years old onwards much faster than its girls counterpart (in red, parameterized by $\beta_2$). This is an evidence that this observation is more likely to arise from a girl record. The sampling of the cluster index \eqref{eq:discreteCC} makes use of the complete data log-likelihood and promotes models involving small deformations. Therefore, the class $I=2$ is more likely as confirmed by Figure \ref{fig:curveSamplingScheme_3} representing the class sampling scheme throughout the $300$ MCMC iterations.

\subsubsection{Template estimation}
Starting with two $\dimF=35$ dimensional random vectors $\hat{\balpha}_{1,0}$ and $\hat{\balpha}_{2,0}$,  the two templates $f_{\hat{\balpha}_{1,1000}}$ and $f_{\hat{\balpha}_{2,1000}}$, displayed in Figure \ref{fig:curves} were obtained after $N=1,000$ iterations of the MCoEM algorithm. Since a limited number of observations are available, each observation is processed several times, drawn at random throughout the iterations. The templates show that the girls reach the pubertal growth spurt earlier (between $11$ and $12$ years) than boys (between $13$ and $14$ years). Moreover, we notice that the boys growth velocity profile features a pre-pubertal dip more pronounced than for the girls.

\subsection{Handwritten digits template extraction}
\label{subsec:5:2}
The algorithm is then applied to a collection of handwritten digits, the US postal database, featuring $N=1,000$ samples for each handwritten digit from $0$ to $9$, each of which consists of a $16 \times 16$ pixel image. The USPS digits data were gathered at the Center of Excellence in Document Analysis and Recognition (CEDAR) at SUNY Buffalo, as part of a project sponsored by the US Postal Service; see \cite{hull:database}. The main difficulty with these data stems from the geometric dispersion \emph{within each class} of digit. Two sources of variability are considered:

\begin{enumerate}[(i)]
\item The first type is assumed meaningful, since intrinsically related to the class of digit, and MCoEM seeks to learn them: the templates. A digit may indeed need more than a single prototype shape to be efficiently model by a mixture of deformable templates. For example, a digit two may be written with or without a loop in the lower left-hand corner and a digit seven may feature an horizontal bar on the diagonal line.
\item The second type is regarded as nuisances resulting from the presentation context and are deemed irrelevant to identify the class of an observation. They consist of small local deformations and global deformations such as a rotations, homotheties and translations. Such nuisances result from the size of the pen used, different handwriting skills, digits being partially censored by the observation window, etc.
\end{enumerate}

\subsubsection{Deformable template model}
An observation $\bY_n$ is a $16\times 16$ matrix, regarded as a $\dimY=256$ dimensional vector, whose coordinates correspond to the photometry of a fixed set of pixels, $(u_{1},\ldots,u_{\dimY})$, such that for all $s\in\{1,\ldots,\dimY\}$, $u_s\in\ooint{-1}{1}\times\ooint{-1}{1}$. The raw database consists of noise-free observations, such that for all $s\in\{1,\ldots,\dimY\}$, $\bY_{n,s}\in\ooint{0}{1}$. To make the problem more challenging, an additive Gaussian noise $W_s=\sigma\epsilon$, where $\sigma=0.2$ and $\epsilon\sim\mathcal{N}(0,1)$, is added to each pixel $\bY_{n,s}$ (see Figure \ref{fig:templates} (a)).

A template $f$ is a function defined on $\Uset=\rset^2$. The dictionary of functions $\{\phi_\ell\}_{\ell=1}^{\dimF}$ is set as Gaussian kernels with $\dimF=256$. The landmark points $\{r_\ell\}_{\ell=1}^{\dimF}$ are regularly spaced in the square $\ooint{-1}{1}\times\ooint{-1}{1}$ and the kernel $\phi_\ell$ is defined as $u\mapsto\phi_\ell(u)=\exp{({\nu}^{-2}(u-r_\ell)^2)}$ with $\nu=0.2$.

Contrary to the growth velocity curve case, where growth profiles feature different scales, the scale dispersion in the measurement space is limited in this database. As a consequence, using a scaling factor $\sca_n$ is not relevant. To each image $\bY_n$ is associated a class $I_n\in\{1,\ldots,C\}$ and a deformation parameter $\defo_n$, such that a template can be geometrically deformed under the action of a function $u\mapsto D(u,\defo_n)$. We consider two complementary types of deformation:
\begin{itemize}
\item A rigid deformation $u\mapsto\rigid{u}{\ups_n}$ where $\ups_n$ parameterizes rotations, homotheties and translations. Indeed, the templates need to be allowed to rotate and to be translated in space, in order to match the observations and in particular those which are partially censored by the observation window. Homotheties allow to zoom in or to zoom out the templates.
     In this case, $\ups_n$ is a $6$-dimensional real vector, $\ups_n=(\varphi_n,\varrho_n,c_n,t_n)$, where $c_n$ is the center of the rotation of angle $\varphi_n$ and of the homotheties having $\varrho_n$ as ratio and $t_n$ is the translation vector. $\rigid{\,\cdot\,}{\ups_n}$ writes for all $u\in\Uset$:
    $$\rigid{u}{\ups_n}=\rotation{\varphi_n}(\varrho_n u+t_n-c_n)+c_n\eqsp,$$
    where $\rotation{\varphi_n}$ is the rotation matrix with angle $\varphi_n$. A Gaussian prior is set on $\ups_n$, with zero mean for the components $(\varphi_n,c_n,t_n)$ and a mean one for $\varrho_n$. The covariance matrix is diagonal with variances set to $0.1$.

\item A smooth small deformation field is used to register locally a template with the observation. It is  parameterized by a $\dimLoc$-dimensional vector $\loc_n=(\loc_{n,1},\ldots,\loc_{n,\dimLoc})$ and writes for all $u\in\Uset$ as
    $$
    \local{u}{\loc_n}=\sum_{k=1}^{\dimLoc}\loc_{n,k}\psi_k(u)\eqsp,
    $$
    where for all $k\in\{1,\ldots,\dimLoc\}$, $\loc_{n,k}\in\rset^{2}$ in order to allow small displacements in the two directions. The smoothness of the deformation is enforced by the choice of functions $\{\psi_k\}_{k=1}^{\dimLoc}$ which belongs to a dictionary of Gaussian kernels defined on $\rset^{2}$ and centered on the landmark points $\{q_k\}_{k=1}^{\dimLoc}$ with identical variance $\sigma_L^{2}$, such that for all $k\in\{1,\ldots,\dimLoc\}$, $\psi_k(u)=\exp{\left(\sigma_{V}^{-2}\|u-q_k\|^{2}\right)}$. In this implementation, we used $\dimLoc=36$ landmark points at the vertices of a regular grid on the square $\ooint{-0.5}{0.5}\times\ooint{-0.5}{0.5}$ and a bandwidth $\sigma_V^2=0.16$. As a consequence the local deformation parameter $\loc_n$ is a $72$-dimensional vector. Similarly to $\ups_n$, conditionally on $I_n=j$, a Gaussian distribution with zero mean and covariance matrix $\Gamma_j$ is assumed for the parameter $\loc_n$. In this implementation, for all $j\in\{1,\ldots,C\}$, $\Gamma_j$ writes $\Gamma_j=\gamma_j^{2}M$ where $M$ is a fixed matrix with ones on the diagonal and $0.2$ on the lower and upper diagonals.
\end{itemize}
Hence, the parameter $\defo_n$ is a $78$-dimensional vector which writes $\defo_n=(\ups_n,\loc_n)$ and belongs to the space $\defoSpace=[0,2\pi]\times\rset^{+}\times\rset^2\times\rset^2\times \rset^{72}$.
\begin{figure}[h]
\includegraphics[scale=0.85]{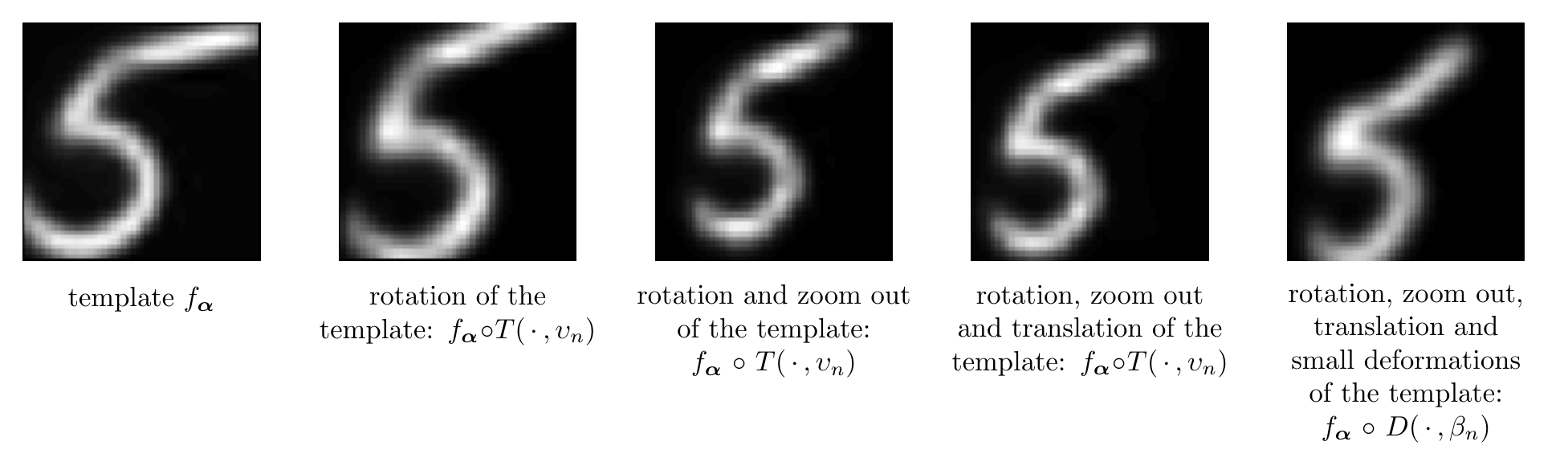}
\caption{Distortion of a template $f_{\balpha}$ under the action of global and local deformations.}
\label{fig:templateDeformations}
\end{figure}

Finally, the deformation model writes in this setting for all $u\in\Uset$ and a parameter $\defo_n\in\defoSpace$ as:
\begin{equation}
\label{eq:defGeoModelDigits}
D(u,\defo_n)=\rotation{\varphi_n}(\varrho_n u+t_n-c_n)+c_n+\sum_{k=1}^{\dimLoc}\loc_{n,k}\psi_k(u)\eqsp.
\end{equation}
It is illustrated with Figure \ref{fig:templateDeformations}.

\newpage
\subsubsection{Parameter estimation}
\label{sec:5:2:2}
\begin{figure}[h]
\begin{tikzpicture}
\draw(0,0) node{\includegraphics[scale=0.6]{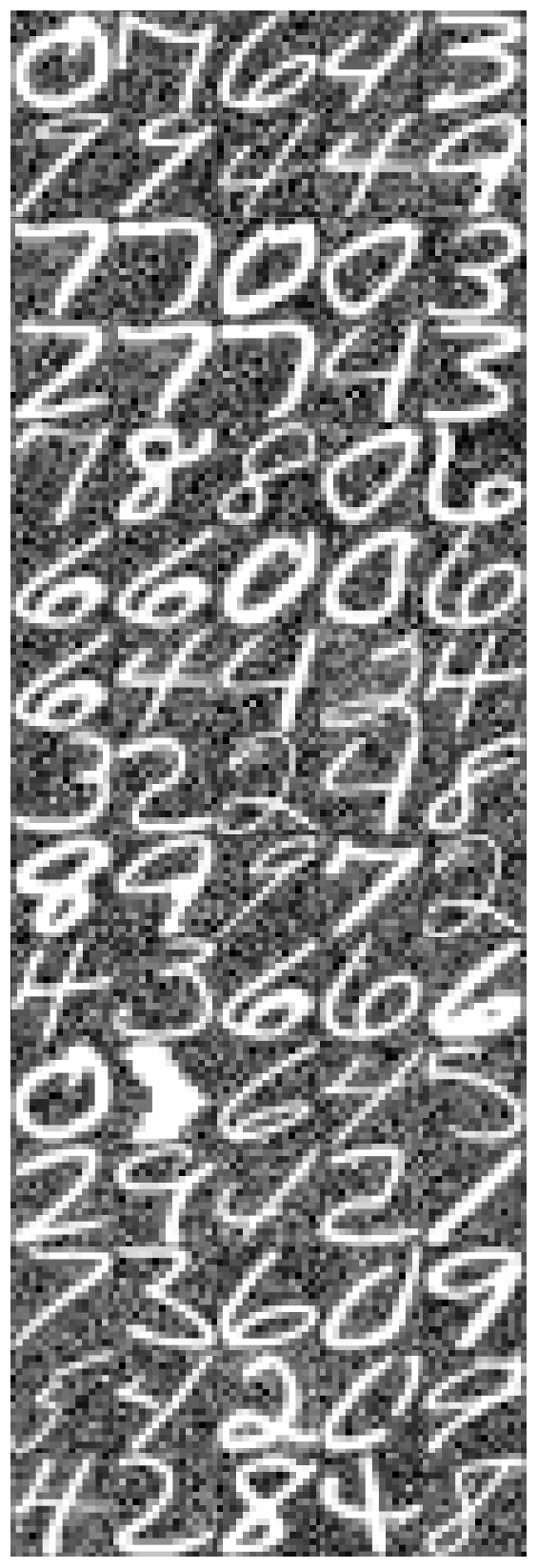}};
\draw(0,-5.8) node{(a) Samples};
\draw(0,-6.3) node{ of handwritten};
\draw(0,-6.8) node{digits};
\draw(6.5,0) node{\includegraphics[scale=1]{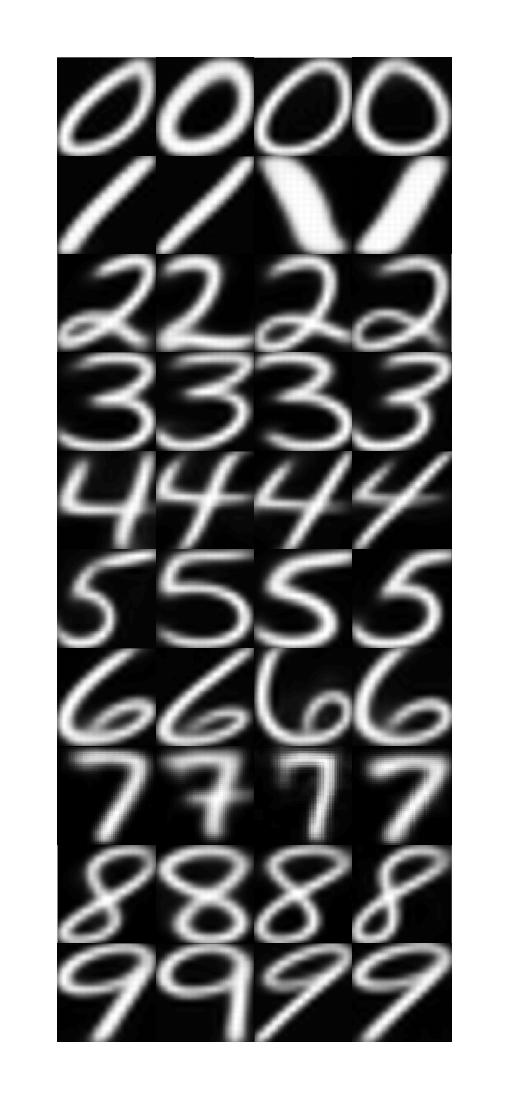}};
\draw(6.5,-5.8) node{(b) Partially supervised scheme};
\draw(12.8,0) node{\includegraphics[scale=1]{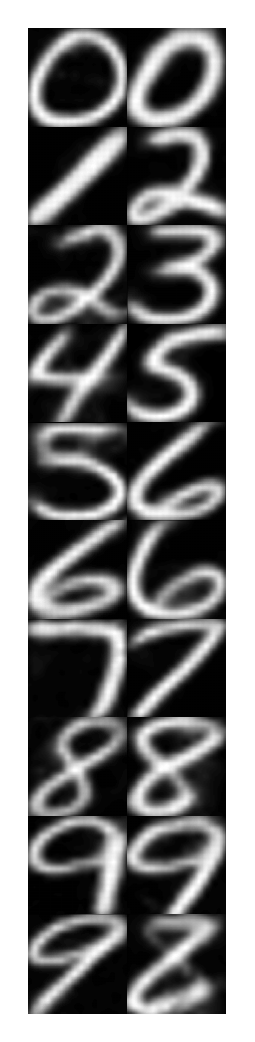}};
\draw(12.8,-5.8) node{(c) Fully unsupervised scheme};
\end{tikzpicture}
\caption{Templates estimated in the two different schemes: (b) partially-supervised, after $N_1=1,000$ MCoEM iterations with $C_1=4$ components for each model and (c) fully-unsupervised, after $N_2=5,000$ MCoEM iterations with $C_2=20$ components. In both setups, MCoEM was applied on handwritten digit images similar to those displayed in (a).}
\label{fig:templates}
\end{figure}

We consider two learning setups:
 \begin{enumerate}
 \item \textbf{Partially-supervised}: the templates are learnt for each digit separately with $C_1=4$ classes through $N_1=1,000$ iterations of the MCoEM. Thus, 10 independent models are learnt and the resulting templates are reported in Figure \ref{fig:templates}-(b). We refer to this approach as partially-supervised since MCoEM deals with images of the same digit (labeled) but assigns each observation to one of the four classes describing this type of digit in an unsupervised fashion.
 \item \textbf{Fully-unsupervised}: the templates are learnt from the dataset containing all the 10 digits (unlabelled), with $C_2=20$ classes and $N_2=5,000$. Thus, only one model is learnt and the resulting templates are illustrated with Figure \ref{fig:templates}-(c).
 \end{enumerate}

The templates obtained in the two settings are similar, even though in Figure \ref{fig:templates}-(c), the algorithm makes use of a class for digits that can hardly be classified in one of the existing mixture component (template in the bottom right corner). In addition, in the fully-unsupervised scheme, the number of classes describing a digit is ruled by the learning algorithm and may not be optimal: for instance a digit two could be described with more than two clusters, whereas three classes for a digit nine are a bit excessive.

Following the guidelines provided in \cite{Cappe:Online}, the sufficient statistics (and consequently the parameters) should be updated for the first time once several observations have been gathered. Indeed, the parameters update step (see Eq. \eqref{eq:mstep_SAOEM}) requires that the sufficient statistics vector check some constraints. In particular $\{\tilde{s}_{n,j,1}\}_{j=1}^{C}$ should be nonzero scalars and $\{\tilde{s}_{n,j,3}\}_{j=1}^{C}$ should be invertible matrices. In practice, these assumptions hold, when the first update happens after $n=50$ MCoEM iterations, the second after $n=75$ and as soon as a new observation is available from $n=100$ onwards. Initialisation of the template parameters can potentially lead to degeneracy if one or more classes are initialised with pathologic parameters. This issue was not encountered in the partially-supervised setup probably because the class sampling is easier, the data being all observations of the same digit. The initial template parameters were thus set randomly. In the fully-unsupervised scheme however, the template parameters were set as the clusters centroid returned by a k-means clustering algorithm (using the Matlab built-in routine) applied to 50 images of the dataset drawn at random. More precisely for all $j\in\{1,\ldots,C\}$, $\hat{\balpha}_{0,j}=(\Phi_{O_{\dimDefo}}^{T}\Phi_{O_{\dimDefo}})^{-1}\Phi_{O_{\dimDefo}}^{T}\bc_j$, where $\Phi_{\defo}$ is defined in Eq. \eqref{eq:matPhi} and $\bc_j$ is k-means cluster $j$ centroid.

\begin{figure}[h]
\begin{tikzpicture}
\draw(-6,1) node{\includegraphics[scale=0.7]{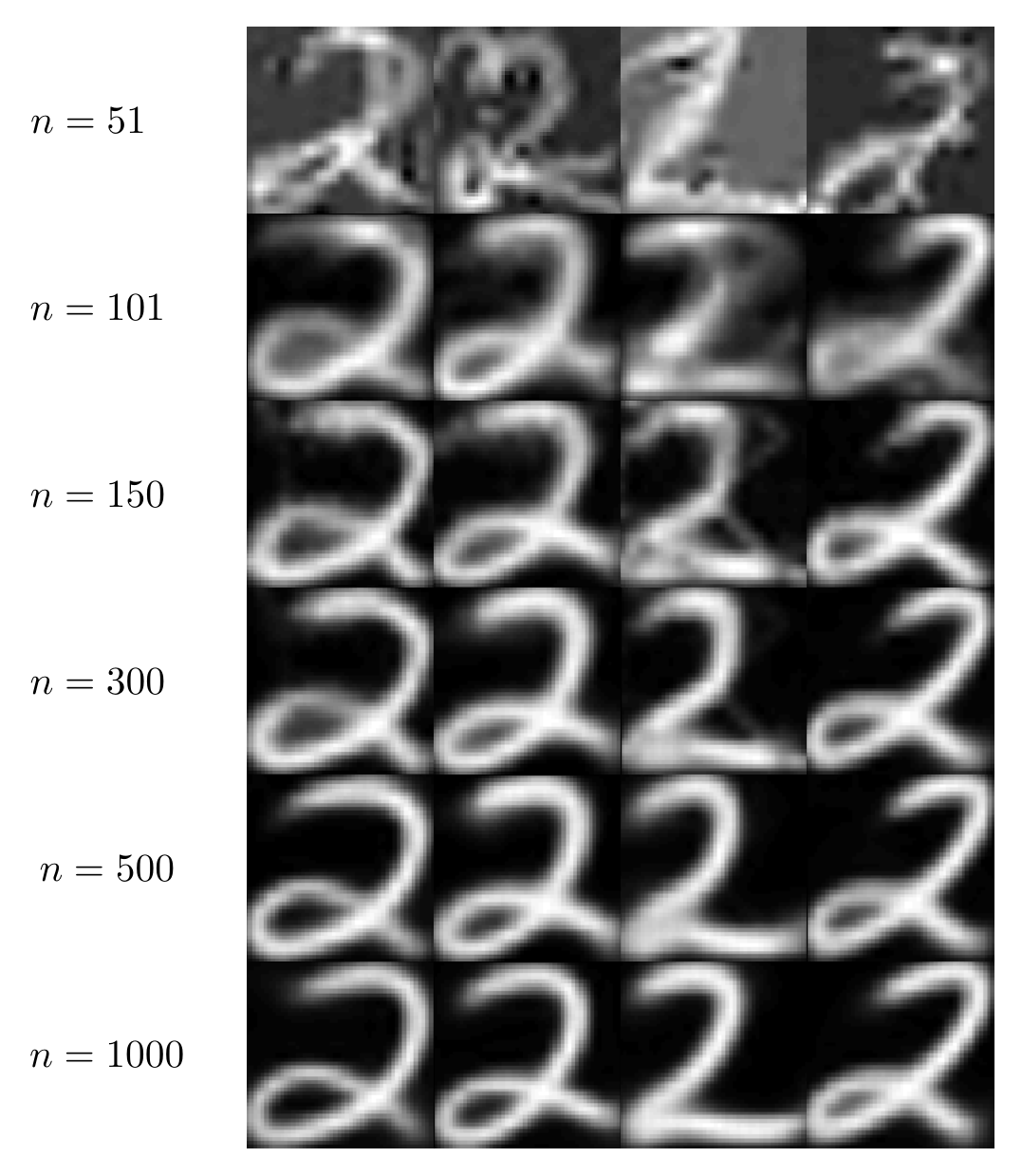}};
\draw(-5,-4) node{(a) Evolution of the templates};
\draw(-5,-4.5) node{$\{f_{\hat{\boldsymbol{\alpha}}_j}\}_{1\leq j\leq C}$};
\draw(2,1) node{\includegraphics[scale=0.62]{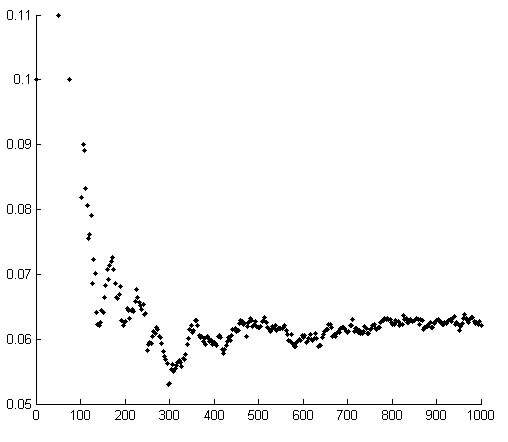}};
\draw(2,-4) node{(b) Evolution of the noise variance};
\draw(2,-4.5) node{estimate $\hat{\sigma}_{n}^{2}$};
\end{tikzpicture}
\caption{Templates extraction and inference}
\label{fig:estimation1}
\end{figure}
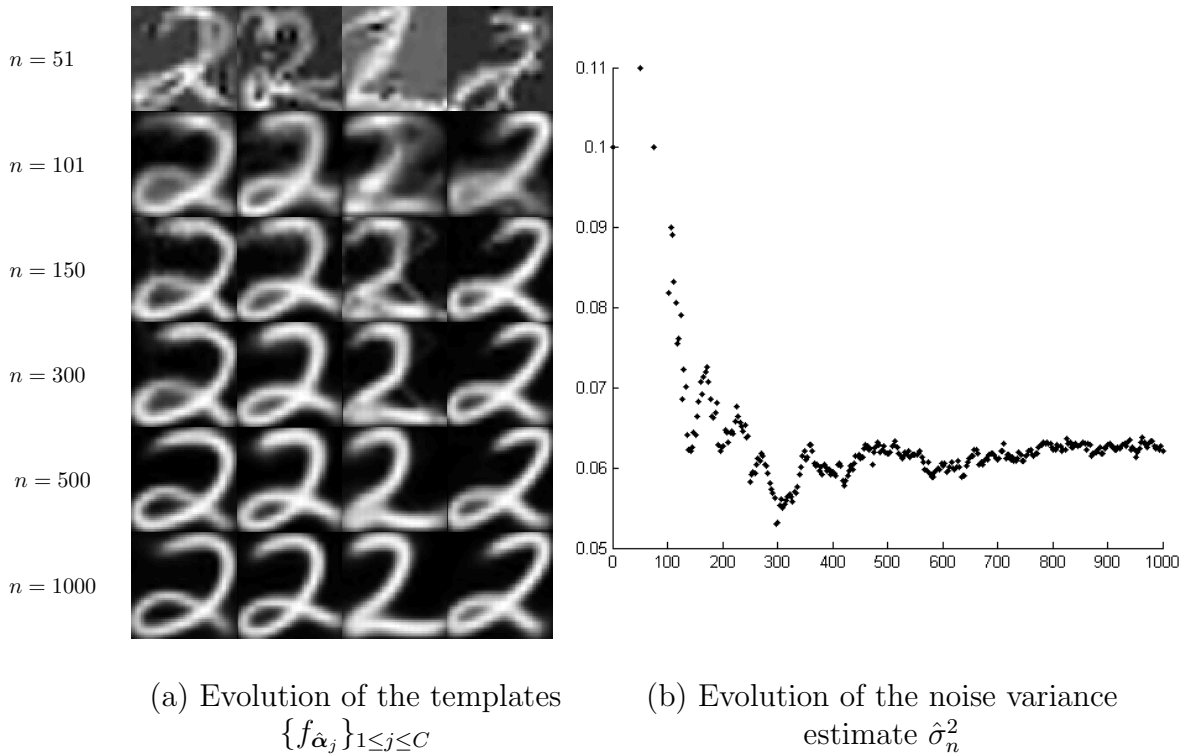

Figure \ref{fig:estimation1} shows the parameters estimate $\{\estim{n}\}_{n=1}^{1000}$ throughout the MCoEM algorithm for the digit two learnt separately with $C=4$ classes (partially-supervised). The functions $\{f_{\balpha_{n,j}}\}_{1\leq j\leq C}$ tends progressively to usual reference shapes and each new observation available enhances the templates estimate (\autoref{fig:estimation1}-(a)).

\subsubsection{Sampling the missing data}
The hidden data $\defo_n=(\ups_n,\loc_n)$, and $I_n$ are simulated with $t_n=200$ iterations if $n\leq 100$ and $t_n=500$ iterations otherwise of the sampling scheme proposed in Section \ref{sec:carlinChib}. This choice of $m_n$ is motivated by the fact that when the templates are not well resolved (which occurs in the early estimates of MCoEM), a rough approximation of the conditional expectation is sufficient. Moreover, a burn-in period of $100$ iterations was applied. Finally, given the high dimension of $\defo_n$, the quasi-Newton optimization methods to estimate $\{\defo_j^{\star}\}_{j=1}^{C}$ in Eq.~\eqref{eq:pseudo_mean} is time-consuming. Therefore, the pseudo-priors parameters are set as the sample mean and covariance matrix derived from $100$ iterations of a random walk targeting the posterior distribution and taking place before the first MCMC iteration.

\begin{figure}
\centering
\includegraphics[scale=0.7]{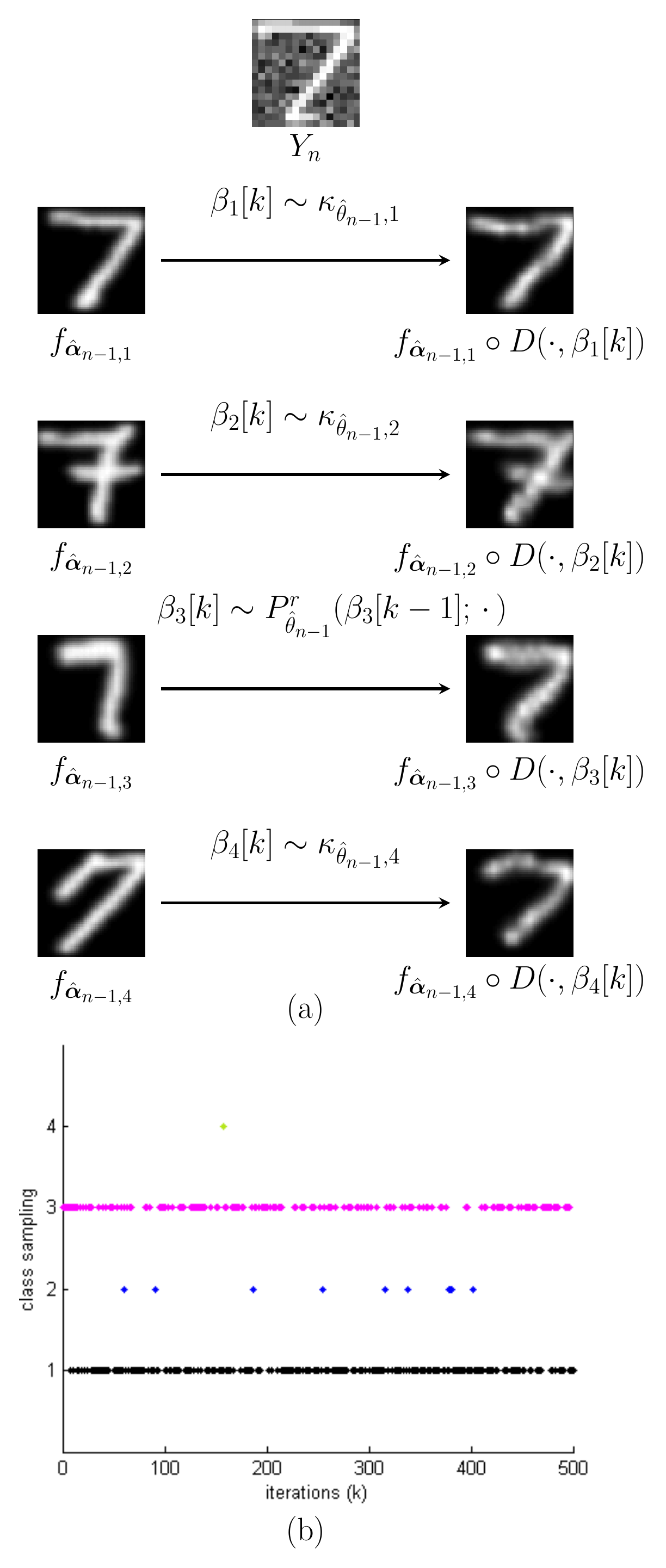}
\caption{Sampling missing data $(I[k],\beta_1[k],\ldots,\beta_4[k])\sim\tilde{\pi}_{\param_n}(\,\cdot\,|\,\bY_n)$ with $n=600$, using the Carlin and Chib approach introduced in Section \ref{sec:carlinChib}. The top panel (a) illustrates the sampling of deformations parameters taking place at the $k=450$-th iteration of the Markov chain. The bottom panel (b) illustrates the class index sampled by the Markov chain. \label{fig:sampling_posterior_digits}}
\end{figure}

Figure \ref{fig:sampling_posterior_digits}-(a) shows a realization of the $k=450$-th iteration of the Markov chain $\tK{n}^{\CC}$ occurring in the $n=600$-th iteration of the MCoEM algorithm (the index $n$ is omitted hereafter). In this scenario, we aim at extracting $C=4$ templates of the digit $7$ in a partially-supervised setting (see Figure \ref{fig:templates}-(b)). Given $I[k-1]=3$, the auxiliary variables $\{\beta_{j}[k]\}_{j\neq 3}$ are sampled from the linking densities $\{\kappa_{\hat{\param}_{n-1},j}\}_{j\neq 3}$, while $\beta_{3}[k]$ is simulated with $r=20$ iterations of a Gaussian increment Random Walk Metropolis-Hastings algorithm, whose variance is adjusted to obtain an overall acceptance rate of $40\%$ (see \cite{Andrieu:IntroMCMC}). Iterating the Metropolis-Hastings kernel $r$ times speeds up the convergence of the chain without changing the stationary distribution. Despite the rough approximation on the pseudo-priors parameters, Figure \ref{fig:sampling_posterior_digits}-(a) shows that the simulated deformations $\beta_{j}[k]$ are consistent with the observation $\bY_{n}$ for each model $j\in\{1,\ldots,C\}$. As a consequence, the Markov chain $\{I[k],\beta_1[k],\ldots,\beta_C[k]\}_{k>0}$ mixes well; see Figure \ref{fig:sampling_posterior_digits}-(b) which displays the class index samples $\{I[k]\}_{k>0}$ throughout the $t_n=500$ MCMC iterations. An animation of the MCMC sampling scheme can be found online at \url{http://mathsci.ucd.ie/~fmaire/MCoEM/carlinChib.html}.

\subsubsection{Comparison with SAEM-MCMC}
For  conciseness, we will write from now on SAEM instead of SAEM-MCMC for the algorithm formalized by \cite{Kuhn:MCMCSAEM} and applied  to perform template estimation in \cite{Allassonniere:Mixture}.

\begin{figure}
\centering
\includegraphics[scale=0.7]{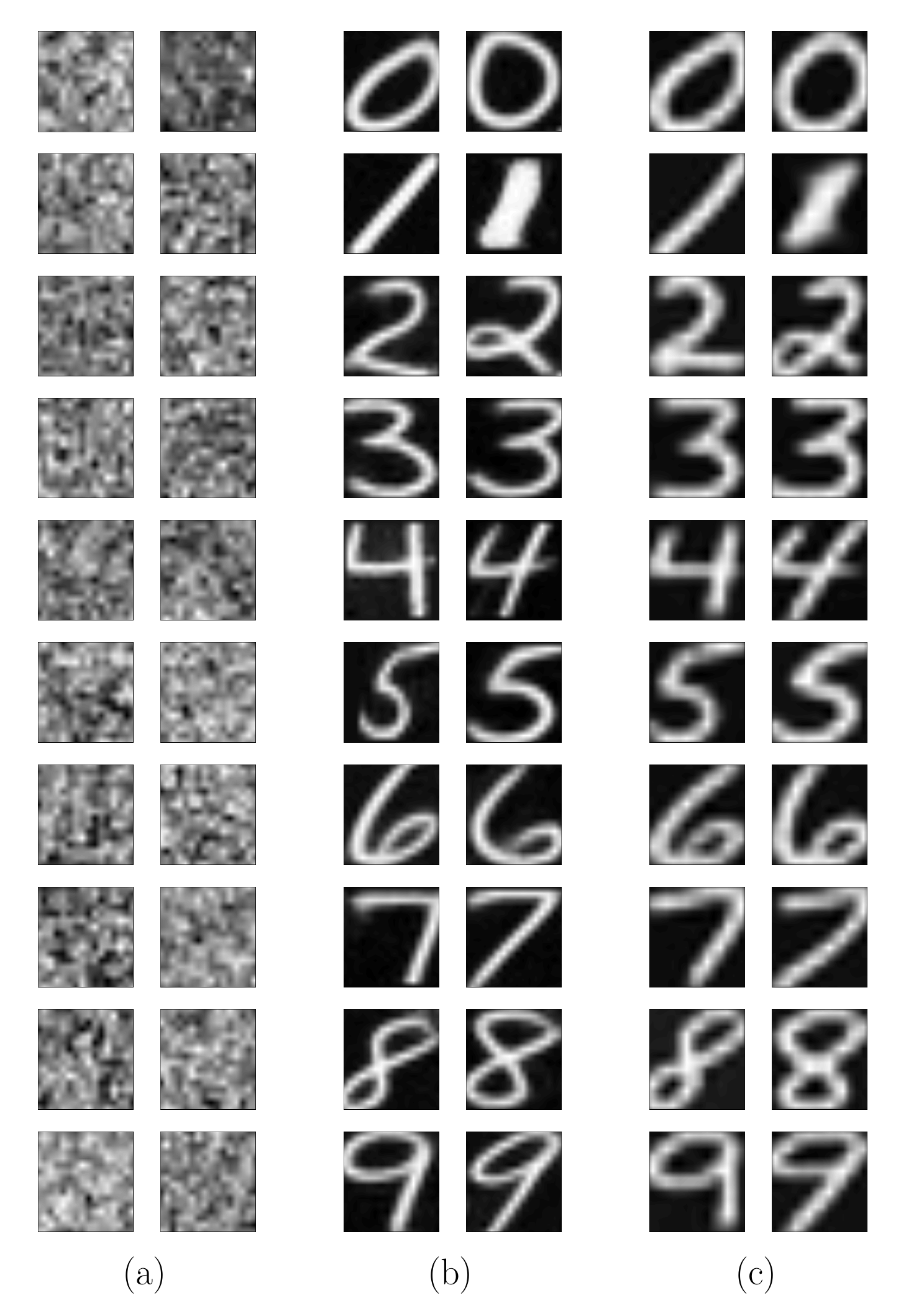}
\caption{Templates extracted by MCoEM (b) and SAEM (c) from the same dataset, consisting of $n=300$ handwritten digit images from each type of digit, in a partially-supervised way and during a 10-hour running time experiment. Each model comprises $C=2$ classes. (a) represents the initial templates drawn at random $\hat{\balpha}_{0,j}\sim\norm(\mathbf{0}_m,M^{-1})$ where $M$ is a square matrix of size $m$ with elements $M_{p,q}=\exp(-\|r_p-r_q\|^2\slash \nu^2)$.\label{fig:templates_mcoem_saem}}
\end{figure}

\begin{figure}
\centering
\includegraphics[scale=0.7]{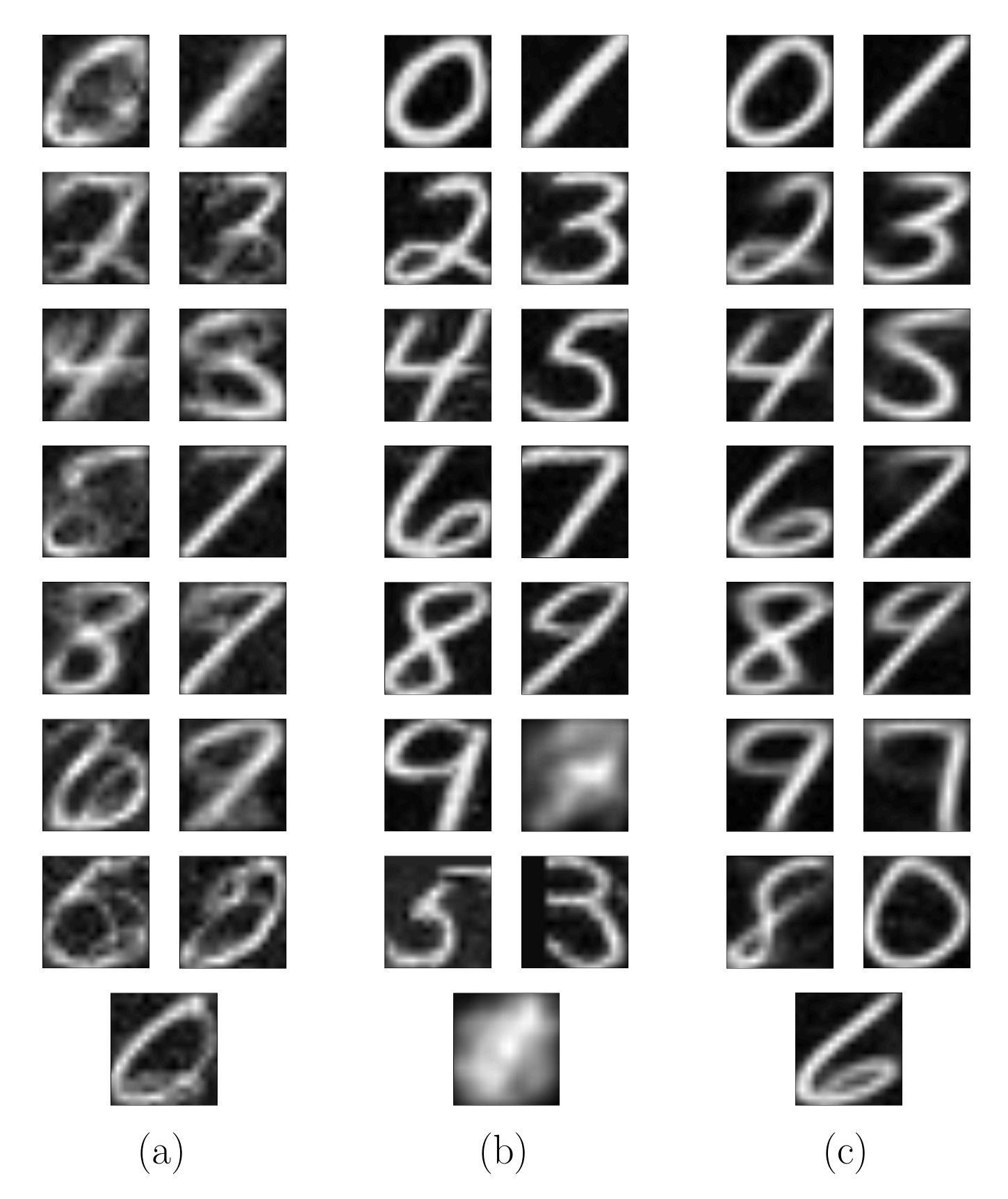}
\caption{Templates extracted by MCoEM (b) and SAEM (c) from the same dataset, consisting of $n=500$ handwritten digit images from each type of digit, in a fully-supervised way and during a 40-hour running time experiment. The model comprises $C=15$ mixture components. (a) represents the initial templates based on k-means clustering applied on 50 random images. See \url{http://mathsci.ucd.ie/~fmaire/MCoEM/templates.html} for an animation. \label{fig:templates_mcoem_saem_unsup}}
\end{figure}

Templates estimated by MCoEM are compared with those obtained by applying SAEM \cite{Allassonniere:Mixture} to the same images, in both setups. In the partially-supervised setup, both algorithms processed the same $n=300$ images for each class of digit, during a 10-hour runtime. In the fully-unsupervised approach, MCoEM and SAEM processed the same $n=500$ images ($50$ images of each digit), during a 40-hour runtime experiment. SAEM is a batch stochastic EM algorithm that processes all the data at each iteration. In the mixture of deformable models context, this means that SAEM has to register each observation with the set of templates estimated at each iteration, whereby a significant computational burden is generated. As a consequence, in a 10-hour running time experiment, SAEM could only perform 23 iterations while MCoEM completed nearly 2,000 iterations. Figures \ref{fig:templates_mcoem_saem} and \ref{fig:templates_mcoem_saem_unsup} report the sets of templates extracted by both methods in the two setups.

In the partially-supervised setup, the two sets of estimated templates show similar features (Fig. \ref{fig:templates_mcoem_saem}), highlighting that in spite of processing the data on the fly, MCoEM yields a similar stability than SAEM. From a qualitative perspective, performing nearly ten times as many iterations than SAEM is beneficial for MCoEM whose templates look much smoother and yield a better resolution. An animation of the template estimation in this setup can be found online at \url{http://mathsci.ucd.ie/~fmaire/MCoEM/templates.html}.

\begin{figure}
\centering
\includegraphics[scale=0.7]{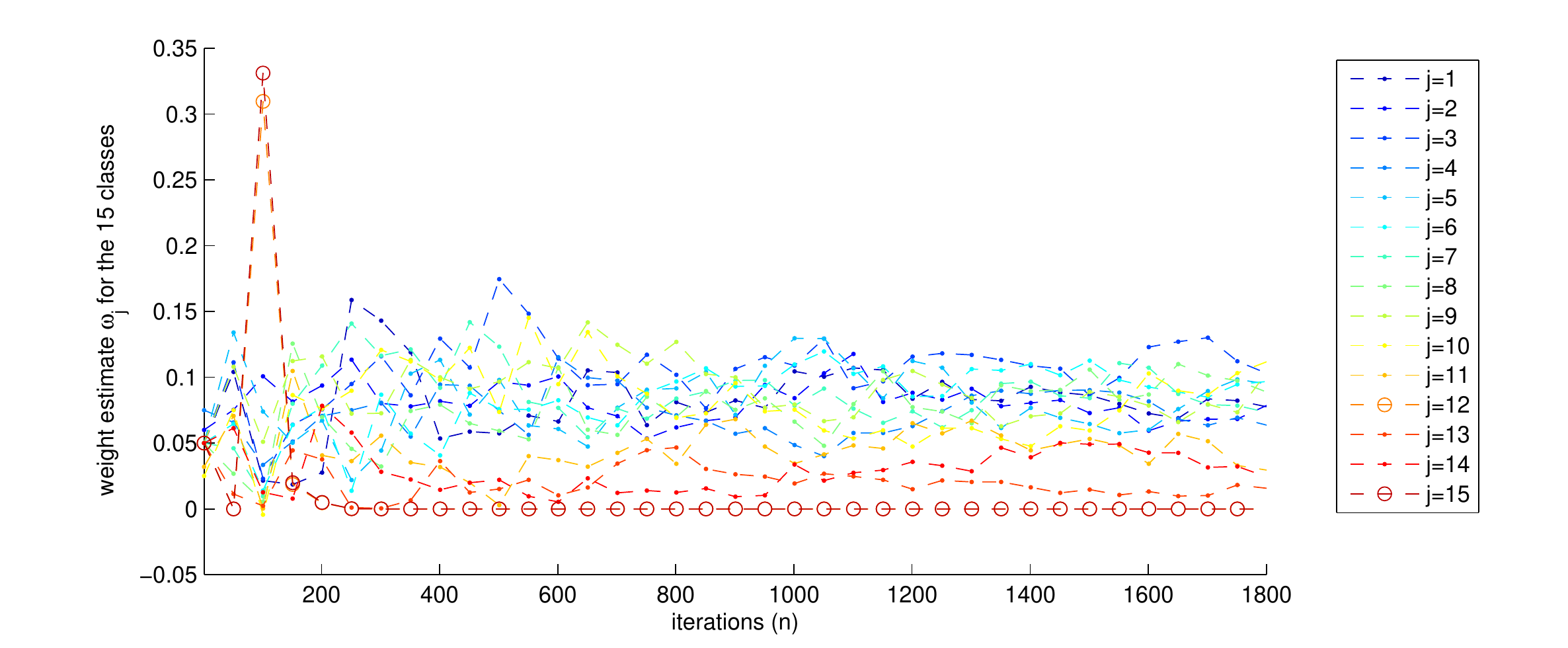}
\caption{Evolution of the weight for each of the $C=15$ classes of the fully-unsupervised mixture of template model inferred by MCoEM. Circled data points corresponds to the two classes whose weight vanishes.\label{fig:weight_est}}
\end{figure}

The templates estimated by MCoEM and SAEM in the fully-unsupervised setup, implemented with $C=15$ components, are reported in Figure \ref{fig:templates_mcoem_saem_unsup}. The first ten templates are consistent for both algorithms while the last five templates differ significantly. On the one hand, MCoEM only makes use of 13 from the 15 available classes. The two remaining classes corresponds to the 12th and 15th templates in the middle column of Figure \ref{fig:templates_mcoem_saem_unsup}. Figure \ref{fig:weight_est} plots the weight evolution for each class as MCoEM moves forward and shows that those two classes have quickly become unused by the algorithm. The first ten classes weight is slightly lower than 1/10 which is in line with the dataset. On the other hand, SAEM maintains the 15 classes alive all throughout the algorithm. In this example, SAEM appears more robust than MCoEM for inferring a mixture model. However, we believe that the stability of MCoEM can be improved by increasing the number of iterations before the first parameter update (only 50 in our simulation), hence avoiding this degeneracy problem. Indeed, from Figure \ref{fig:weight_est} it is clear that those two classes have been left empty after the first 50 iterations, paving the way to the pathological effect observed at the next updates.

\section{Classification}
\label{sec:classification}
When considering real-time classification applications, the MCoEM methodology may prove more adequate than SAEM: indeed as soon as the first estimate of $\param$ is available, a classifier can be implemented. Of course, the rate of correct classification is expected to improve upon random guessing as soon as the templates take shape. Both algorithms produce a sequence of parameter estimates. However, since iterations of MCoEM and SAEM have different complexity, we consider $\hat{\theta}_t$, the parameter estimate after a runtime of $t$ time units, as a fair way to compare both methods.

Learning parameters of the mixture of deformable models \eqref{eq:mixtModelVect} allows to classify labeled observations $\{(\btY_1,V_1),\ldots,(\btY_N,V_N)\}$ gathered in a testing dataset. There is no overlap between those testing observations and the data $\{\bY_1,\ldots,\bY_n\}$ processed by the algorithms during the learning phase. Let $\rho_t$ be the \textit{live} error rate at time $t$, defined as the empirical rate of uncorrect classification (based on $N=1,000$ testing observations) obtained using the parameters estimated by the algorithms at time $t$:
$$
\rho_t=\frac{1}{N}\sum_{k=1}^N\mathds{1}_{\hat{V}_{k,t}\neq V_k}\,,
$$
where $\hat{V}_{k,t}$ is the class of digit assigned to $\btY_k$ returned by the classifier using the estimate $\param_t$. In this Section, we compare the live error rate on the handwritten digits example such that $V_k\in\{0,\ldots,9\}$ (see Section \ref{subsec:5:2}) based on estimates from MCoEM (processing a new observation at each iteration), SAEM-50 and SAEM-300, \ie SAEM using $n=50$ and $n=300$ learning observations respectively. Both learning setups partially-supervised and fully-unsupervised are considered.

\subsection{Partially-supervised learning}
In this approach, each type of digit $v\in\{0,\ldots,9\}$ is described at time $t$ by a set of parameters $(\hat{\theta}_{1,t}^{(v)},\ldots,\hat{\theta}_{C,t}^{(v)})$. We used $C=2$ classes per digit in this implementation. The following unnormalized probabilities
\begin{equation}
\label{eq:classif1}
\text{for all }v\in\{0,\ldots,9\},\quad \pi_v(\btY_k,\hat{\param}_{t})=\sum_{i=1}^{C}\esp_{\hat{\param}_{i,t}^{(v)}}\left[g_{\theta}(\btY_k\,|\,I_k,\bX_k)\,|\,\btY_k, I_k=i\right]\,,
\end{equation}
are calculated and the guess $\hat{V}_{k,t}$ is defined as
\begin{equation}
\label{eq:classif1_V}
\hat{V}_{k,t}(\hat{\theta}_t)=\arg\max_{v\in\{0,\ldots,9\}}\pi_v(\btY_k,\hat{\theta}_{t})\,.
\end{equation}
The conditional expectation in \eqref{eq:classif1_V} is intractable and approximated by the sample mean of a Metropolis--Hastings Markov chain targeting the posterior distribution of $\bX_k$, $\pi(\,\cdot\,|\,\bY_k,I_k=i)$.

\begin{figure}
\centering
\includegraphics[scale=.7]{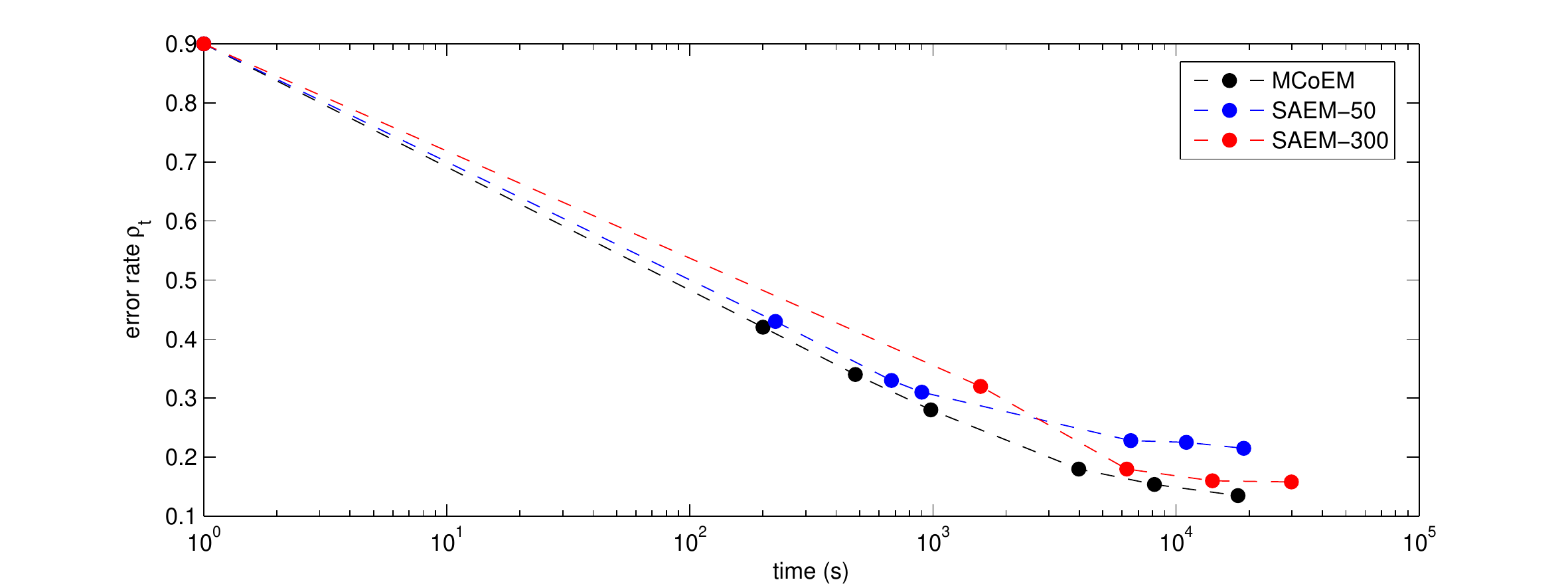}
\caption{Live error rate for MCoEM, SAEM-50 and SAEM-300, applied in the partially-supervised setup. For each algorithm, a ball at time $t$ represents the error rate at time $t$. $\rho_t$ is obtained by comparing the estimated class $\hat{V}_{k,t}$ with the label $V_k$ for the $N=1,000$ testing data. $\hat{V}_{k,t}$ is returned by the classifier making use of the knowledge acquired by the algorithm up to time $t$.  Dashed lines are used for readability only and do not convey any error rate outside the balls.\label{fig:classif_partially}}
\end{figure}

Figure \ref{fig:classif_partially} reports the live error rate for the three algorithms applied in the partially-supervised setup. For each algorithm the error rate at $t=0$ is $\rho_0=.9$ since the initial template are uninformative (see Figure \ref{fig:templates_mcoem_saem}-(a)) The second left-most ball corresponds to the error rate using the first estimate produced by each algorithm. For each algorithm, the time per iteration is reported in Table \ref{tab:classRunTime}. Note that for numerical stability, MCoEM first parameter update occurred after 10 iterations, the second after 15 iterations and at each iteration from the 20-th iteration onwards. SAEM yields a large error rate when learning from only $n=50$ observations. It is significantly reduced when using $n=300$ observations but this improvement comes at the price of a prohibitively large computational cost. The first estimate produced by SAEM-300 is available after $t_1=1,570$ s. At this time, MCoEM has already performed nearly 80 iterations and exhibits a significantly lower live error rate: approximately $\rho_{t_1}=.24$ for MCOEM and $\rho_{t_1}=.34$ for SAEM-300. MCoEM yields a successful tradeoff between SAEM with limited $n$ allowing quick estimates but poor error rate and SAEM with larger $n$ allowing lower error rate but a slower estimation. In addition, since MCoEM makes use of new data at each iteration, its live error rate is expected to keep reducing while SAEM's error rate, using a fixed dataset, seems to flatten once convergence of the parameter is reached.

\begin{table}
\centering
\caption{CPU time of an iteration of MCoEM and SAEM with $n=50$ and $n=300$ and $C=2$ mixture components.\label{tab:classRunTime}}
\vspace{.2cm}
\begin{tabular}{c|c|c|c}
 & MCoEM & SAEM-50 & SAEM-300  \\
\hline
CPU time / iteration (s) & 20 & 225 & 1,570\\
\end{tabular}
\end{table}

\subsection{Fully-unsupervised learning}
\label{sec:classification:2}

Since in this learning setup, an object (\eg a digit nine) may be described by several templates (3 in the simulation of Figure \ref{fig:templates}-(c) and 2 in that of Figure \ref{fig:templates_mcoem_saem_unsup}-(b)), an intermediate layer of classes is required. Based on this observation, an external agent must specify the mapping $M:\Iset\to\{0,1,\ldots,9\}$ that links each class designed by MCoEM/SAEM to the object it describes. Classification is then carried out as in the semi-supervised setup. More precisely, given the estimate $\hat{\theta}_t$, the following unnormalized probabilities
\begin{equation}
\label{eq:classif2}
\text{for all }v\in\{0,\ldots,9\},\quad \pi_v(\btY_k,\hat{\param}_t)=\sum_{i\in M(v)}\esp_{\hat{\param}_{i,t}}\left[g_{\theta}(\btY_k\,|\,I_k,\bX_k)\,|\,\btY_k,I_k=i\right]\,,
\end{equation}
are approximated by an MCMC estimate and the guess $\hat{V}_{k,t}$ is derived as in \eqref{eq:classif1_V}.

\begin{figure}
\centering
\includegraphics[scale=.7]{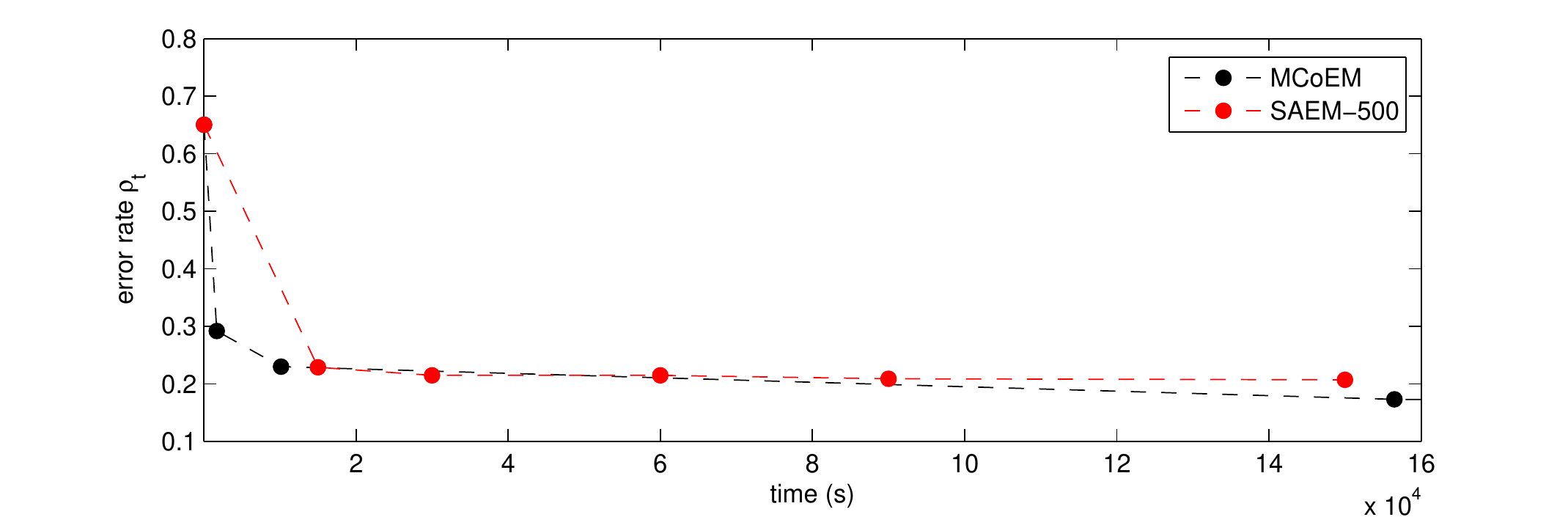}
\caption{Live error rate for MCoEM and SAEM-500, applied in the fully-unsupervised setup. Dashed lines are used for readability only and do not convey any error rate outside the balls. \label{fig:classif_unsup}}
\end{figure}

Figure \ref{fig:classif_unsup} reports the live error rate of MCoEM and SAEM-500 in the fully-unsupervised setup.
At time $t=0$, the error rate $\rho_0$ is .65 and not .9 as in the previous setup. This is because the initial templates are derived from k-means centroids based on the same $n=50$ data (see Section \ref{sec:5:2:2}) and are thus no longer non-informative. SAEM is clearly penalized by processing $n=500$ observations and estimating $C=15$ classes of parameters and it nearly takes 5 hours of computation to get the first SAEM's estimate. Interestingly, SAEM's first estimate is nearly as "good", in the error rate sense, as the MCoEM estimate obtained after 5 hours. Nevertheless, using MCoEM offers a practitioner the possibility to classify much quicker new observations.

\begin{table}
\centering
\caption{CPU time of an iteration of MCoEM and SAEM $n=500$ and $C=15$ mixture components.\label{tab:classRunTime_unsup}}
\vspace{.2cm}
\begin{tabular}{c|c|c}
 & MCoEM & SAEM-500   \\
\hline
CPU time / iteration (s) & 170 & 17,570\\
\end{tabular}
\end{table}
\section{Discussion}
\label{sec:conclusion}
We have proposed a statistical framework to perform sequential and unsupervised inference in a deformable template model, with application to curve synchronization and shape extraction and registration. It makes use of the Monte Carlo online EM algorithm (MCoEM), derived from \cite{Cappe:Online} and a novel MCMC sampling method, based on the Carlin and Chib sampler \cite{Carlin:Bayesian}, allowing to simulate the unsamplable joint distribution of the cluster index and deformation parameters. The method has been applied successfully to extract reference templates from several data sets featuring high time/geometric dispersion.

Our work was primarily motivated by the computational gain arising when processing one observation at a time. Indeed, when the missing data is a large vector and many observations are available, stochastic batch EM algorithms such as SAEM \cite{Delyon:ConvSAEM} are prohibitively slow for practical use. This has been illustrated with the classification problem (Section \ref{sec:classification:2}) in which SAEM's error rate after nearly 5 hours of computation is still at the initial level. In comparison, it took MCoEM less than 20 minutes to reach less than half the initial error rate. In this perspective, MCoEM can be regarded as a linearization of stochastic batch EM algorithms, which can be particularly appealing in a Big Data context.

In terms of implementation, the main concern when inferring a mixture model with MCoEM is class degeneracy. To mitigate this risk, two points have been discussed. First, a particular care should be brought to the way initial parameters are set and especially the templates. We have suggested to use k-means clustering on a limited set of observations to initiate the templates. Second, the number of EM iterations between the first parameter updates should be large enough in order to assign at least one observation to each class. Adaptive implementations have not been considered but could yield an automated update schedule.

The handwritten digit example studied in this paper shows that MCoEM seems to inherit SAEM's asymptotic behaviour. Indeed, (i) qualitatively, the template shapes extracted by both algorithms are similar and (ii) quantitatively, the error rates are comparable. This result calls for further investigation as a theoretical framework is yet to be developed to establish the convergence of MCoEM. Both SAEM and online EM proofs of convergence relies on stochastic approximation theory arguments. However those proofs cannot be straightforwardly extended to MCoEM since it combines two approximations: one on the conditional expectation (which is in SAEM) and the other one on the data generating process (which is in the online EM). We therefore leave this as a future work. An interesting question is to assess to what extend the convergence rate of the online EM \cite{Cappe:Online}, known to be optimal, is degraded when replacing the expectation of the sufficient statistics by an unbiased estimate.
\section*{Acknowledgments}
\noindent This work has been supported by the ONERA, the French Aerospace Lab and the DGA, the French Procurement Agency.

\bibliographystyle{elsarticle-num}
\bibliography{mml16}

\end{document}